\documentclass[journal, 12pt,onecolumn, draftclsnofoot]{IEEEtran}
\usepackage{ragged2e}
\usepackage{floatrow}
\usepackage{upgreek}
\usepackage{gensymb}
\usepackage{enumerate}
\usepackage{amsthm}
\usepackage[utf8]{inputenc}
\usepackage{soul}            
\usepackage[english]{babel}
\theoremstyle{remark}
\newtheorem{theorem}{Theorem}
\newtheorem{lemma}{Lemma}

\usepackage{multicol}
\usepackage{epsfig}
\usepackage{epstopdf}
\usepackage{float}
\usepackage{verbatim}
\usepackage{perpage}
\MakeSorted{figure}
\usepackage{array}
\usepackage{graphicx}
\usepackage{amsfonts}
\usepackage{ulem}
\usepackage{amssymb}

\usepackage{enumitem}
\usepackage[english]{babel}
\usepackage[utf8]{inputenc}
\usepackage{algorithmic}
\usepackage[dvipsnames]{xcolor}


\usepackage{mathtools}

\usepackage{hyperref}
\hypersetup{linktocpage} 
  \hypersetup{
    colorlinks,
    citecolor=black,
    filecolor=black,
    linkcolor=black,
    urlcolor=black
}

\usepackage{amsmath}
\usepackage{cite}
\DeclareMathAlphabet\mathbfcal{OMS}{cmsy}{b}{n}
\ifCLASSINFOpdf
\else
\fi
\usepackage{array}
\usepackage{fixltx2e}

\usepackage[linesnumbered,ruled,vlined]{algorithm2e}
\usepackage{environ}

\newcounter{parentAlgoLine}
\SetKwBlock{BEGIN}{}{}

\makeatletter
\NewEnviron{substeps}{%
  \refstepcounter{AlgoLine}
  \protected@edef\theparentequation{\theequation}%
  \setcounter{parentAlgoLine}{\value{AlgoLine}}%
  \setcounter{AlgoLine}{0}%
  \BEGIN{\BODY}%
  \setcounter{AlgoLine}{\value{parentAlgoLine}}%
  \ignorespacesafterend
}
\makeatother
\usepackage[font=footnotesize,labelfont=bf, figurename=Fig.]{caption} 
\usepackage{subcaption}
\usepackage{fix-cm}
\usepackage{floatrow}
\newfloatcommand{capbtabbox}{table}[][\FBwidth]

\usepackage{blindtext}

\usepackage{etoolbox} 
\newtoggle{SINGLE_COL}
\toggletrue{SINGLE_COL}     

\newtoggle{BIG_EQUATION}
\toggletrue{BIG_EQUATION}  

\iftoggle{SINGLE_COL}{
\setlength{\abovedisplayskip}{1pt}
\setlength{\belowdisplayskip}{1pt}
}{}

\newcommand{\Hmat}{\mathbf{H}}

\newcommand{\CNcal}{\mathcal{CN}}

\newcommand{\hv}{\mathbf{h}}

\begin{document}

\title{Variational Learning Algorithms For Channel Estimation in RIS-assisted mmWave Systems}
\author{\IEEEauthorblockN{Milind Nakul, Anupama Rajoriya, and Rohit Budhiraja}}

\maketitle
\vspace{-2cm}
\begin{abstract}
We consider the problem of estimating the channel in reconfigurable intelligent surface (RIS) assisted millimeter wave (mmWave) systems. We propose two variational expectation maximization (VEM) based algorithms for channel estimation in RIS-aided wireless systems. The first algorithm is a structured mean field based sparse Bayesian learning (SM-SBL) algorithm that exploits the doubly-structured sparsity as well as the individual sparsity of the elements of the channel. To exploit the sparsities we propose a column-wise coupled Gaussian prior. We next design the factorized mean field based algorithm based on the prior we propose. This algorithm called the factorized mean field SBL (FM-SBL) algorithm addresses the time complexities of the SM-SBL algorithm without sacrificing channel estimation accuracy. We show using extensive numerical investigations that the i) proposed SM-SBL and FM-SBL algorithms outperform several existing algorithms and ii) FM-SBL has lower time complexity compared to the SM-SBL algorithm.

\end{abstract}


\IEEEpeerreviewmaketitle
\section{Introduction}
Reconfigurable intelligent surface (RIS) is a promising technology to assist mmWave systems in achieving high spectral and energy efficiency. An RIS consists of a large number of passive reflecting elements which phase shifts the incident signal, and reflects it passively. Each of the RIS elements reflects the incident signal independently, and these elements can be configured such that the received signal combines coherently, thus improving the received signal power \cite{gamp}. This is especially useful in cases when the direct path between the user (UE) and the base station (BS) is blocked by obstructions \cite{gamp}.

There has been extensive research in the area of joint optimization of the beamformer  and the RIS reflection coefficient matrix \cite{LS1,opti1,opti2}. These works assume that the accurate channel state information (CSI) is available. Accurate CSI hence becomes extremely important for deployment of RIS-aided wireless systems. However, the channel estimation in these systems is difficult due to the large number of passive reflecting elements \cite{gamp}. The end-to-end channel in RIS-assisted wireless systems, the \textit{cascaded channel}, is a product of two separate components, the UE-RIS and RIS-BS channel. The cascaded virual angular domain (VAD) channel, due to the limited number of scatterers around the BS and RIS exhibits sparsity.

References \cite{active-elements,LS2,LS1,Binary1} have estimated the cascaded channel of the RIS-aided wireless systems. Taha \textit{et al.} in \cite{active-elements} estimated the channel by including active elements in the RIS which have signal processing capability. These active elements receive the signal and help in estimating the UE-RIS and RIS-BS channels. This design uses compressive sensing tools to contruct channels of all the RIS elements from the channels observed only at the active elements. The installation of active elements with signal processing capability in RIS is, however, costly and may not be suitable for many applications \cite{gamp}. Mishra \textit{et al.} in \cite{LS1} have proposed a channel estimation algorithm for multiple input single output systems that are assisted by RIS. It uses a binary reflection least squares estimator to estimate the cascaded channel matrix. For channel estimation only one of the elements is turned on at a time. The BS then estimates the cascaded channel from the UE to the BS. The ON/OFF reflection modes need to be implemented for each individual passive RIS element. The ON/OFF techniques leads to a sub-optimal solution \cite{LS2}. Jensen \textit{et al.} in \cite{LS2} proposed minimum variance unbiased estimator to estimate the cascaded channel. This design does not require any prior knowledge about the channel and employs a least squares estimation technique. He \textit{et al.} in \cite{Binary1} proposed a two step channel estimation protocol for RIS-aided massive MIMO systems. The first step of the algorithm estimates the UE-RIS channel using matrix factorization techniques. The second step estimates the RIS-BS channel using matrix completion. It assumes that the reflection coefficient matrix of the RIS is binary and requires it to be sparse for accurate channel estimation. The ON/OFF switching of the passive RIS elements in \cite{LS1,Binary1} is costly as this requires separate amplitude control of each RIS element\cite{ON/OFF}. The channel estimation techniques considered in \cite{active-elements,LS1,LS2,Binary1} do not exploit the inherent sparsity of the cascaded VAD channel, hence, they require large training overhead for accurate channel estimation \cite{gamp}.

The cascaded VAD channels for different UEs have a doubly-structured sparsity with completely common non-zero rows and partially-common non-zero columns. The authors in \cite{gamp} exploited the sparsity of the cascaded channel to express the downlink channel estimation problem in RIS-aided mmWave systems as compressed signal recovery. It then uses the orthogonal matching pursuit (OMP) and generalized approximate message passing (GAMP) to solve the compressed sensing problem. The GAMP algorithm, in order to incorporate the sparsity structure, assumed a Gaussian mixture prior on the channel. The OMP based algorithm is, however, greedy and requires the knowledge of the sparsity degree of the signal to be recovered. Ruan \textit{et al.} in \cite{amp} proposed a two step protocol by expressing the channel estimation as a compressed sensing recovery problem. It requires the deployment of a testing site in order to aid with the channel estimation process. In the offline phase the test site T1 transmits pilot signals to BS for estimating the cascaded T1-RIS-BS channel to obtain the RIS-BS channel. In the online phase  T1 first transmits pilot signals to UEs for estimating the cascaded T1-RIS-UE channel to obtain the RIS-UE channel. It then uses the vector approximate message passing (VAMP) algorithm to solve the compressed sensing recovery problem. To capture the sparsity structure of the cascaded VAD channel, a Bernoulli-Gaussian family of prior is assumed. The AMP based algorithms in \cite{gamp,amp} do not exhaustively exploit the complete sparsity of the cascaded VAD channel in the RIS-aided wireless systems.

 The authors in \cite{ris} introduced and leveraged the doubly-structured sparsity of the angular domain cascaded channels associated with different UEs. They proposed a doubly-structured orthogonal matching pursuit (DS-OMP) based algorithm. This method jointly estimates the row and the column supports of all the UEs. This reduces the pilot overhead of the cascaded channel estimation. The authors in \cite{TS-OMP} have proposed another aspect in the sparsity structure of the cascaded VAD channel of different UEs. They have proposed that in addition to the doubly-structured sparsity the cascaded VAD channel also has an additional column index shift which is common across all the UEs. They proposed that for each UE the non-zero rows of the cascaded VAD channel has a shift relative to each other which is independent of the UE. After compensating for this shift all the non-zero rows of each UE have the same column support. Based on this they have proposed a three-step OMP (TS-OMP) algorithm for channel estimation. The algorithms \cite{TS-OMP, ris} require sparsity information of cascaded VAD channel, i.e., the number of non-zero rows and columns, for channel estimation. These are acquired with extremely high computational complexity for large number of BS antennas or RIS elements \cite{amp}.

References \cite{gamp, ris, TS-OMP, amp} which exploit sparsity of the cascaded channel in RIS-assisted wireless systems, propose the channel estimation problem as compressed signal recovery. We can thus apply various compressed sensing algorithms in order to obtain accurate channel estimates. The sparse bayesian learning (SBL) framework introduces a suitable prior over the sparse channel to be recovered and estimates the channel using probablistic methods \cite{tipping}. The choice of the prior here is very important as it must accurately capture the sparsity structure of the channel. These SBL algorithms are, however, iterative in nature and suffer from time complexity issues \cite{fmf-sbl}. Reference \cite{fmf-sbl} proposes a faster version of the SBL algorithm using variational inference (VI). These methods have not been applied for channel estimation in RIS-aided mmWave systems.


To summarize, the existing literature for RIS-assisted wireless communication systems has not yet designed a VI based channel estimation technique which is able to exhaustively exploit the entire sparsity structure of the cascaded VAD channel. VI proposes the inference problem in channel estimation as an optimization problem. This allows us to approximate even intractable posterior distibutions and also allows us to use different simpler forms of the approximating distribution for faster convergence \cite{fmf-sbl}. In this work we estimate the channel of the RIS-assisted mmWave systems. We exploit the doubly-structured sparsity of the cascaded VAD channel in order to reduce the channel estimation pilot overhead. Our main contributions can be summarized as follows, \newline
1) We first propose a column-wise coupled prior which accurately captures the sparsity structure of the cascaded VAD channel in RIS-aided systems. The proposed prior, along with the joint estimation of the cascaded VAD channel of all UEs helps in reduncing the pilot overhead of channel estimation.\newline
2) We develop a structured mean field sparse Bayesian learning (SM-SBL) algorithm for accurate channel estimation in RIS-assisted mmWave systems. The developed algorithm accurately estimates the cascaded VAD channel along with hyperparameters of the proposed prior. The SM-SBL algorithm maximizes a bound on the log marginal likelihood of the model called the evidence lower bound (ELBO). The proposed algorithm has much lower training overhead than the existing algorithms \cite{ris,TS-OMP}.\newline
3)The SM-SBL algorithm  suffers from time complexity issues due to the computational cost of each iterative step in the algorithm. We then propose an algorithm which uses VI along with the Lipschtiz inequality to get a faster version of the algorithm, referred to as the factorized mean field sparse Bayesian learning (FM-SBL) algorithm. The Lipschitz inequality is used to bound the ELBO along with the factorized posterior assumption of VI to come up with updates that do not require matrix inversions, this speeds up the mean and variance computation steps.\newline
4) We numerically show that the proposed SM-SBL and FM-SBL algorithms, have a lower normalized mean squared error (NMSE) and lower pilot overhead than the existing algorithms\cite{sbl,ris}. The proposed algorithms also have low normalized support error rate (NSER) which shows that they accurately estimate the true support of the cascaded VAD channel. We plot the Kullback–Leibler (KL) divergence to show that the proposed FM-SBL algorithm estimates a distribution which is very close to the distribution estimated by the SM-SBL algorithm. This plot also helps us in validating that the Lipschitz inequality in the FM-SBL algorithm does not degrade the performance. We finally show that the proposed FM-SBL algorithm has much lower runtime than the SM-SBL algorithm.

\textbf{Notations:}
The $m$th element of vector $\boldsymbol{y}$ is denoted as {$y_m$}. The superscripts, $( \cdot)^{T}$, $(\cdot)^*$, $(\cdot)^H$ are transpose, conjugate and conjugate transpose operations. The symbols  {${x}_{m,n}$, $\mathbf{x}_m^T$, $\mathbf{x}_{n}$, } denote the $(m,n)$th element, $m$th row and $n$th column of matrix $\mathbf{X}$, respectively.  
 The notations $\mathbb{C}^{K\times M}$, $\mathbb{R}_+^{K\times M}$, $\left \{0,1 \right \}^{K\times M} $ represent complex, positive real valued and binary matrices respectively, of dimension $K\times M$. The symbol $\mathbf{I}_M$, $\mathbf{1}_{M }$ and $\mathbf{0}_{M}$ denote $M\times M$ identity matrix,  $M \times 1$ all-one and all-zero vectors, respectively. The notations $\mathbb{I}(\cdot)$ and $\phi$ denote the indicator function and the null set, respectively.  The notations $\|\cdot\|_2$ and $\|\cdot\|_F$ denote the $\ell_2$ and Frobenious norm of a vector and matrix, respectively. The statistical expectation with respect to a distribution $p(\cdot)$ is denoted as $\mathbb{E}_{p(\cdot)}(\cdot)$. The notation $\text{diag}(\mathbf{y})$ represents a diagonal matrix with the elements  of $\mathbf{y}$ on its diagonal. 
  The notation $\mathcal{CN}(x\vert a,b)$ implies that the random variable $x$ has complex Gaussian distribution with mean $a$ and variance $b$. The support set of a vector/matrix is denoted as $\text{supp}(\cdot)$.

\vspace{-0.5cm}
\section{System model}\label{sysmodel}
We consider an RIS-assisted mmWave system with an $M$-antenna BS and $K$ single-antenna UEs. The RIS is degined as a uniform planar array (UPA) with  $N_1$ horizontal and $N_2$ vertical reconfigurable elements.  The total number of RIS elements is $N=N_1 \times N_2$. Each RIS element is passive , and has no signal processing capability. It only shifts the phase of the incident signal \cite{ris}.  The uplink channel from UE to RIS is  denoted as $\mathbf{h}_{rk}\in \mathbb{C}^{N \times 1}$, while that of between the BS to RS is denoted as $\mathbf{G} \in \mathbb{C}^{M\times N}$. 
This work focuses on estimating the cascaded uplink UE-RIS-BS channel at the BS. The direct channel between the UE and BS can be easily estimated using the conventional estimation schemes\cite{ris}, and is not considered in this work.

Each UE, similar to \cite{ris,TS-OMP},  transmits a known orthogonal pilot over $Q$ time slots to the BS to estimate the uplink cascaded channel. Let $s_{kq}$ be the pilot signal transmitted by the $k$th UE in the $q$th time slot, for $q = 1 \text { to } Q$. The pilot signal received at the BS from the $k$th UE in the $q$th time slot  after removing the contribution of the direct channel is expressed as \cite{ris}:
\begin{align}\label{eq:measurement_vector}
    \mathbf{y}_{kq} & = \mathbf{G}  \text{diag} (\boldsymbol{\theta}_{q})\mathbf{h}_{rk}s_{kq} + \mathbf{w}_{kq}
\stackrel{(a)}{=} \mathbf{G}  \text{diag} (\mathbf{h}_{rk})\boldsymbol{\theta}_{q}s_{kq} + \mathbf{w}_{kq}.
\end{align}
Here $\boldsymbol{\theta}_{q} = [\theta_{q,1},\theta_{q,2},\dots,\theta_{q,N}]^{T}\in \mathbb{R}^{N \times 1}$ is the RIS reflection vector, with  $\theta_{q,n}$, for $n= 1 \text { to } N$,  being the reflection coefficient of the $n$th RIS element in the $q$th time slot. Equality in (a) is obtained by interchanging $\boldsymbol{\theta}_q$ and $\mathbf{h}_{rk}$, as $\text{diag} (\boldsymbol{\theta}_q)$ is a diagonal matrix.  The vector $\mathbf{w}_{kq}$, with pdf $\mathcal{CN}(0,\sigma^{2}\mathbf{I}_{M})$, is the additive white Gaussian noise(AWGN) at the BS. 
The objective is to estimate the cascaded channel $\mathbf{H}^{k} \triangleq \mathbf{G}$diag$(\mathbf{h}_{rk})$. To achieve it, we re-express \eqref{eq:measurement_vector} as follows:
\begin{align}\label{eq:y_kq}
    \mathbf{y}_{kq} = \mathbf{H}^{k}\boldsymbol{\theta}_{q}s_{kq} + \mathbf{w}_{kq}.
\end{align}
The $Q$ time-slot pilot transmissions are concatenated, for the sake of mathematical convenience, as  $\mathbf{Y}^{k} = [\mathbf{y}_{k1},\dots,\mathbf{y}_{kQ}] \in \mathbb{C}^{M \times Q}$. Since the objective is to estimate the cascaded  UE-RIS-BS channel for the given RIS reflection coefficients,  Eq. \eqref{eq:y_kq} is simplified by assuming that $s_{kq}=1$, $\forall$ $k$ UEs and $q$ slots, as follows:
\begin{align}\label{eq:Y_k}
    \mathbf{Y}^{k} = \mathbf{H}^{k}\boldsymbol{\Theta} + \mathbf{W}^{k},
\end{align}
where $\boldsymbol{\Theta} = [\boldsymbol{\theta}_{1},\dots,\boldsymbol{\theta}_{Q}] \in \mathbb{C}^{N \times M}$ and $\mathbf{W}^{k} = [\mathbf{w}_{k,1},\dots,\mathbf{w}_{kQ}] \in \mathbb{C}^{M \times Q}$.
\vspace{-0.6cm}
\subsection{Virtual angular domain representation of the cascaded UE-RIS-BS channel}
The cascaded channel for the $k$th UE $\mathbf{H}^{k}$ in \eqref{eq:y_kq} is next equivalently represented in the virtual angular domain (VAD) \cite{ris}, where it exhibits sparsity due to the limited number of scatterers around the BS and RIS. This as shown later, will aid its estimation. The $\mathbf{H}^{k}$ can be represented in VAD as $\mathbf{H}^{k}$ \cite{ris}:
\begin{align}\label{eq:angular}
    \mathbf{H}^{k} = \mathbf{U}_{M}(\mathbf{\tilde{H}}^{k})^H\mathbf{U}_{N}^{T}.
\end{align}
Here $\mathbf{\tilde{H}}^{k}$ is the cascaded VAD channel for the $k$th UE, $\mathbf{U}_{M}$ and $\mathbf{U}_{N}$ are the $M \times M$ and $N \times N$ unitary dictionary matrices at the BS and RIS, respectively. Each column of $\mathbf{U}_{M}$ represents the array steering vector corresponding to an angle of arrival (AoA) at the BS \cite{TS-OMP}. Similarly, the each column of $\mathbf{U}_N$ represents the array steering vectors corresponding to a particular angle of departure (AoD) at the RIS. These dictionary unitary matrices uniformly sample the AoA and AoD angles in $[-\pi/2,\pi/2]$, and accurately capture the true physical channel\cite{cp-sbl}. By substituting the VAD representation from \eqref{eq:angular} into (\ref{eq:Y_k}), we have
\begin{align}\label{eq:Y_k_angular}
    \mathbf{Y}^{k} = \mathbf{U}_{M}(\mathbf{\tilde{H}}^{k})^H\mathbf{U}_{N}^{T}\boldsymbol{\Theta} + \mathbf{W}^{k}.
\end{align}
Pre-multiplying (\ref{eq:Y_k_angular}) by $\mathbf{U}_{M}^{H}$, and by taking the conjugate transpose, we have
\begin{align}\label{eq:system_model}
    \tilde{\mathbf{Y}}^{k} = \tilde{\boldsymbol{\Theta}}\tilde{\mathbf{H}}^{k} + \tilde{\mathbf{W}}^{k}.
\end{align}
Here $\tilde{\mathbf{Y}}^{k} = (\mathbf{U}_{M}^{H}\mathbf{Y}^{k})^H \in \mathbb{C}^{Q \times M}$ is the effective received pilot matrix, $\tilde{\boldsymbol{\Theta}} = (\mathbf{U}_{N}^{T}\boldsymbol{\Theta})^{H} \in \mathbb{C}^{Q \times N}$ is the effective phase matrix, and $\tilde{\mathbf{W}}^{k}=(\mathbf{U}_{M}^{H}\mathbf{W}^{k})^H \in \mathbb{C}^{Q \times M}$ is the effective noise matrix.

A joint system model for all the UEs using \eqref{eq:system_model} can be expressed as follows
\begin{align}\label{eq:system_model_stacked}
	\tilde{\mathbf{Y}} = \tilde{\boldsymbol{\Theta}}\tilde{\mathbf{H}} + \tilde{\mathbf{W}}.
\end{align}
Here $\tilde{\mathbf{Y}} = [(\tilde{\mathbf{Y}}^{1})^{T}, \hdots , (\tilde{\mathbf{Y}}^{K})^{T}]^{T} \in \mathbb{C}^{QK \times M}$ is the effective joint receive pilot matrix, $\tilde{\boldsymbol{\Theta}} \in \mathbb{C}^{QK \times MK}$ is the joint effective phase matrix and $\tilde{\mathbf{H}} \in \mathbb{C}^{NK \times M}$ is the joint effective channel. If the $m$th column of $\tilde{\mathbf{H}}$ and $\tilde{\mathbf{H}}^{k}$ are represented as $\tilde{\mathbf{h}}_{m}$ and $\tilde{\mathbf{h}}_{m}^k$ respectively, then 
\begin{align}
	\tilde{\mathbf{h}}_{m} = [(\tilde{\mathbf{h}}_{m}^1)^{T},\hdots,(\tilde{\mathbf{h}}_{m}^K)^T]^T \in \mathbb{C}^{NK \times 1}.
\end{align}

We next take a closer look at the cascaded channel $\mathbf{H}^{k}$  and its corresponding VAD representation $\tilde{\mathbf{H}}^{k}$ from \eqref{eq:angular}. To perform this investigation, the Saleh-Valenzuela channel model from \cite{ris} is used to represent the BS-RIS channel $\mathbf{G}$ as follows \cite{ris}:
\begin{align}\label{eq:channelG}
    \mathbf{G} = \sqrt{\frac{MN}{L_C}}\sum_{l=1}^{L_C}\alpha_{l}^{G}\mathbf{b}(\vartheta_{l}^{G_{r}},\psi_{l}^{G_{r}})\mathbf{a}(\vartheta_{l}^{G_{t}},\psi_{l}^{G_{t}})^{T}.
\end{align}
Here $L_C$ is the number of scatterers between the RIS and  BS, $\alpha_{l}^{G}$ is the complex gain of the $l$th path, $\vartheta_{l}^{G_{r}}(\text{resp. } \psi_{l}^{G_{r}})$ are the azimuth (resp. elevation) angle at the BS, and $\vartheta_{l}^{G_{t}}(\text{resp. }\psi_{l}^{G_{t}})$ are the azimuth (resp. elevation) angle at the RIS for the $l$th path. The RIS-$k$th UE channel $\mathbf{h}_{rk}$ can similarly be represented as follows \cite{ris}:
\begin{align}\label{eq:channel_hrk}
    \mathbf{h}_{rk} = \sqrt{\frac{N}{L_R^k}}\sum_{l_{2}=1}^{L_R^k}\alpha_{l_{2}}^{rk}\mathbf{a}(\vartheta_{l_{2}}^{rk},\psi_{l_{2}}^{rk})^{T}.
\end{align}
The term $L_R^k$ is the number of paths between the $k$th UE and the RIS, $\alpha_{l_{2}}^{rk}$ is the complex gain of the $l_{2}$th path, and $\vartheta_{l_{2}}^{rk}$, and $\psi_{l_{2}}^{rk}$ are respectively the the azimuth and elevation angles at the RIS for the $l_{2}$th path. The vectors $\mathbf{b}(\vartheta,\psi)\in \mathbb{C}^{M \times 1}$ and $\mathbf{a}(\vartheta,\psi) \in \mathbb{C}^{N\times1}$ represent the normalized array steering vectors at the BS and the RIS respectively. For a UPA of size $N = N_{1} \times N_{2}$, the normalized array steering vector at the RIS $\mathbf{a}(\vartheta,\psi)$ can be represented as follows \cite{ris}:
\begin{align}\label{eq:yt}
    \mathbf{a}(\vartheta,\psi) = \frac{1}{\sqrt{N}}\left[e^{-j2\pi d sin(\vartheta)\cos(\psi)\mathbf{n}_{1}/\lambda}\right]\otimes\left[e^{-j2\pi d \sin(\vartheta)\mathbf{n}_{2}/\lambda}\right].
\end{align}
The normalized array steering vector at the BS $\mathbf{b}(\vartheta,\psi)$ can similarly be represented as \cite{TS-OMP}:
\begin{align}
	\mathbf{b}(\vartheta,\psi) = \frac{1}{\sqrt{M}}\left[e^{-j2\pi d sin(\vartheta)\cos(\psi)\mathbf{m}_{1}/\lambda}\right]\otimes\left[e^{-j2\pi d \sin(\vartheta)\mathbf{m}_{2}/\lambda}\right].
\end{align}
Here $\mathbf{n}_{1}=[0,1,\dots,N_{1}-1]^{T}$, $\mathbf{n}_{2}=[0,1,\dots,N_{2}-1]^{T}$, $\mathbf{m}_{1}=[0,1,\dots,M_{1}-1]^{T}$, $\mathbf{m}_{2}=[0,1,\dots,M_{2}-1]^{T}$ , $\lambda$ is the carrier wavelength, and $d$ is the antenna spacing. 
\vspace{-0.6cm}
\subsection{Doubly-structured sparsity of the angular cascaded channel}
\begin{figure}[htbp]
	\centering
	\includegraphics[scale=0.45]{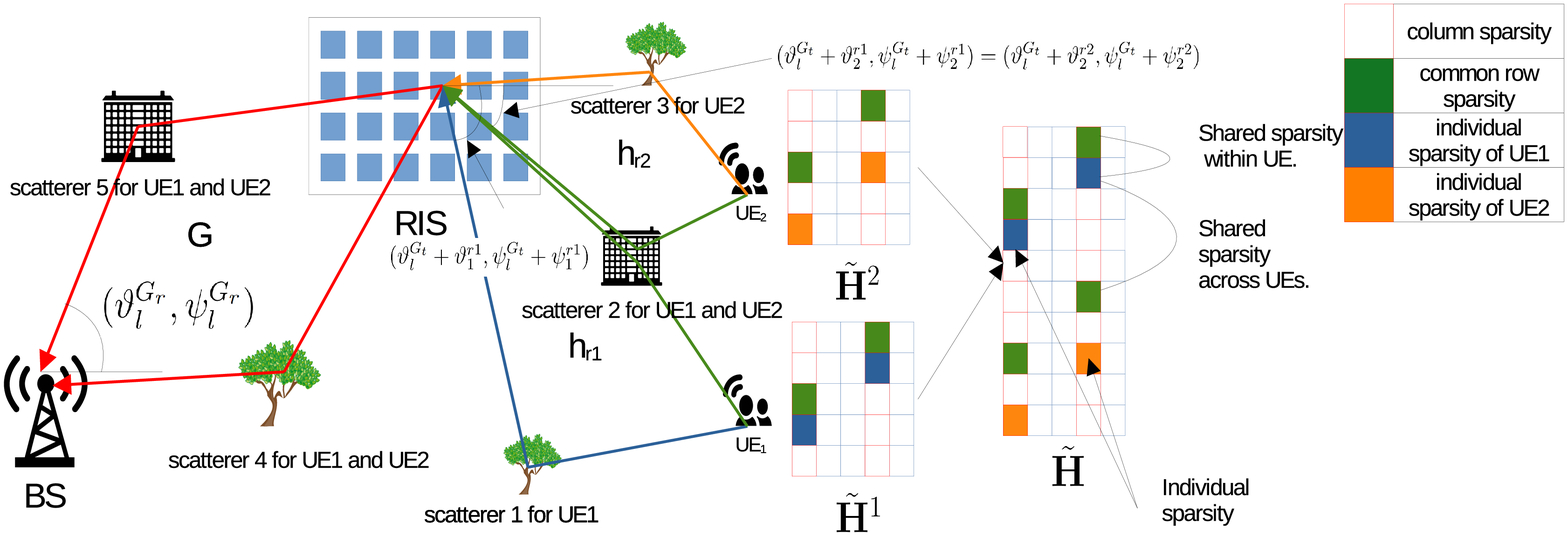}
	\caption{Doubly-structured sparsity in the virtual angular domain (VAD) cascaded UE-RIS-BS channels.}
	\label{fig:double_sparsity}
\end{figure}

From (\ref{eq:channelG}) and (\ref{eq:channel_hrk}), the cascaded VAD channel  in (\ref{eq:angular}) can be re-expressed as \cite{ris}:
\begin{align}\label{eq:angular_cascaded}
    \tilde{\mathbf{H}}^{k} &= \mathbf{U}_N^{T}(\mathbf{G}\text{diag}(\mathbf{h}_{rk}))^H\mathbf{U}_M =  \sqrt{\frac{MN^{2}}{L_CL_R^k}}\sum_{l=1}^{L_C} \sum_{l_{2}=1}^{L_R^k} (\alpha_{l}^{G})^*(\alpha_{l_{2}}^{rk})^* 
    \tilde{\mathbf{a}}(\vartheta_{l}^{G_{t}}+\vartheta_{l_{2}}^{rk},\psi_{l}^{G_{t}}+\psi_{l_{2}}^{rk})^{T}\tilde{\mathbf{b}}(\vartheta_{l}^{G_{r}},\psi_{l}^{G_{r}}).
\end{align}
Here $\tilde{\mathbf{b}}(\vartheta,\psi) = \mathbf{b}(\vartheta,\psi)^{H}\mathbf{U}_{M}$ and $\tilde{\mathbf{a}}(\vartheta,\psi) = \mathbf{a}(\vartheta,\psi)^{H}\mathbf{U}_{N}$.
Both $\tilde{\mathbf{b}}(\vartheta,\psi)$ and $\tilde{\mathbf{a}}(\vartheta,\psi)$ have only one non-zero element which lies at the position of array steering vector in the direction $(\vartheta,\psi)$ in $\mathbf{U}_{M}$ and $\mathbf{U}_{N}$ \cite{ris}. This happens because each column of $\mathbf{U}_{M}$ and $\mathbf{U}_{N}$ is given by the array steering vector corresponding to one specific angle, i.e.,
\begin{align}
	\mathbf{U}_{M} &= \left[ \mathbf{b}(\vartheta_1^{G_r},\psi_1^{G_r}), \hdots,\mathbf{b}(\vartheta_{M_1}^{G_r},\psi_1^{G_r}),\mathbf{b}(\vartheta_1^{G_r},\psi_2^{G_r}),\hdots, \mathbf{b}(\vartheta_{M_1}^{G_r},\psi_{M_2}^{G_r})    \right] \in \mathbb{C}^{M \times M}, \\ \nonumber
	\mathbf{U}_{N} &= \left[ \mathbf{a}(\vartheta_1^{G_t},\psi_1^{G_t}),\hdots,\mathbf{a}(\vartheta_{N_1}^{G_t},\psi_1^{G_t}),\mathbf{a}(\vartheta_1^{G_t},\psi_2^{G_t}),\hdots \mathbf{a}(\vartheta_{N_1}^{G_t},\psi_{N_2}^{G_t}) \right] \in \mathbb{C}^{N \times N}.
	\end{align}
 When $\mathbf{U}_M$ and $\mathbf{U}_N$ are multiplied with $\mathbf{b}(\vartheta,\psi)^H$ and $\mathbf{a}(\vartheta,\psi)^H$ respectively, only that element of $\tilde{\mathbf{b}}(\vartheta,\psi)$ and $\tilde{\mathbf{a}}(\vartheta,\psi)$, is non-zero, which corresponds to the angle $(\vartheta,\psi)$. The non-zero element of $\tilde{\mathbf{b}}(\vartheta,\psi)$ corresponds to a particular scatterer between the BS and RIS and results in the column sparsity of $\tilde{\mathbf{H}}^{k}$. In Fig. \ref{fig:double_sparsity}, the scatterer 5 between the RIS and BS causes the element of $\tilde{\mathbf{b}}$ corresponding to $(\vartheta_l^{G_r},\psi_l^{G_r})$ AoA at the BS to be non-zero. This makes, as shown in Fig. \ref{fig:double_sparsity}, one column of $\tilde{\mathbf{H}}^1$ and $\tilde{\mathbf{H}}^2$ non-zero. Similarly,  the non-zero element of $\tilde{\mathbf{a}}(\vartheta,\psi)$ corresponds to the scatterer between the UE and RIS and results in row sparsity of $\tilde{\mathbf{H}}^{k}$. In Fig. \ref{fig:double_sparsity} the scatterer 1 for UE1 causes $\tilde{\mathbf{a}}(\vartheta_{l}^{G_{t}}+\vartheta_{1}^{r1},\psi_{l}^{G_{t}}+\psi_{1}^{r1})$ to be non-zero. This makes one element inside the $l$th non-zero column of $\tilde{\mathbf{H}}^1$ and $\tilde{\mathbf{H}}^2$ non-zero. The matrix $\tilde{\mathbf{H}}^{k}$, consequently, has two different kinds of sparsities, which are discussed below.

\subsubsection{Common column sparsity} The decomposition of cascaded VAD channel $\tilde{\mathbf{H}}^{k}$ in (\ref{eq:angular_cascaded}) shows that only $L_C$ of its columns are non-zero, which correspond to the $L_C$ scatterers between  the BS and RIS. As seen from \eqref{eq:angular_cascaded}, the non-zero column indices depend on the azimuth and elevation angle at the BS denoted by $(\vartheta_{l}^{G_{r}},\psi_{l}^{G_{r}})$. We also observe from \eqref{eq:y_kq} that the channel $\mathbf{G}$ is common to all the $K$ UEs. The angles  $\{ (\vartheta_{l}^{G_{r}},\psi_{l}^{G_{r}}) \} _{l=1}^{L_C}$ are, thus, same for all the UEs. The non-zero column support of $\tilde{\mathbf{H}}^{k}$, denoted as $\Omega_{c}^{k}$ is common for all $K$ UEs i.e., 
$     \Omega_{c}^{1} = \Omega_{c}^{2} = \dots =\Omega_{c}^{K} = \Omega_{c}$. 
Here $\Omega_{c}$ is the non-zero common column support of $\tilde{\mathbf{H}}^{k}$ for all $k \in [1,\hdots,K]$ UEs. It is shown for two different UEs in Fig. \ref{fig:double_sparsity}, and is given as $\Omega_{c} = \{1,4\}$. The two common non-zero columns are a result of scatterer 4 and scatterer 5 between the RIS and BS. These common column supports are denoted by the red outlines. Since this support is common for the UEs, the shared sparsity across them can be exploited by jointly estimating their channel. 

\subsubsection{Partially-common row sparsity} As discussed earlier, in each non-zero column, only a few rows are non-zero due to the limted number of scatterers around the RIS. As seen from (\ref{eq:angular_cascaded}), the indices of these non-zero rows is decided by the azimuth and elevation angle at the RIS denoted as $\{(\vartheta_{l}^{G_{t}}+\vartheta_{l_{2}}^{rk},\psi_{l}^{G_{t}}+\psi_{l_{2}}^{rk})\}_{l_2=1}^{L_R^k}$. In each non-zero column,  there are thus $L_R^k$ non-zero rows, which correspond to the number of scatterers between the RIS and UE. Some of these scatterers, as shown by scatterer 2 in Fig. \ref{fig:double_sparsity},  will be common for different UEs. These common scatterers will lead to $\{\mathbf{h}_{rk}\}_{k=1}^K$ having partially-common paths with the same angle at the RIS as shown by the green lines in Fig. \ref{fig:double_sparsity}. If the number of partially-common paths between the UE and RIS is denoted as $L_R$, with $L_R \leq L_R^k$, then for each $l$ we can find that, there exists $L_R$ common paths for the cascaded VAD channel between the $K$ UEs and BS, $\{\tilde{\mathbf{H}}_k\}_{k=1}^K$ \cite{ris}.
This means that for each non-zero column $l$ ($l=1,2,\hdots, L_C$), the channels of all $K$ UEs i.e., $\{\tilde{\mathbf{H}}_k\}_{k=1}^K$ share $L_R$ common non-zero rows. The common non-zeros rows in each column across the UEs is denoted by the green boxes in Fig. \ref{fig:double_sparsity} If we denote the non-zero row entries of the $l$th column and $k$th UE as $(\Omega_{r}^{l})^k$, then
    \begin{align}\label{eq:common_column_support}
       \bigcap_{k=1}^{K} (\Omega_{r}^{l})^k =  \Omega_{r}^{l}.
    \end{align}
    Here $\Omega_{r}^{l}$ denotes the partially-common row support for the $l$th column $\forall \; k$.
We show this partially-common row sparsity of cascaded VAD channels in Fig. \ref{fig:double_sparsity}   for two different UEs where the number of paths between the i) BS and RIS is $L_C=2$; and ii)  UE and RIS for each UE is $L_R^k=2$. We see that there the one shared scatterer between the RIS and the UEs is scatterer 2. Out of the $L_R^k$ paths, thus $L_R=1$ path is common for both the UEs to the RIS. This implies that for each UE ,there are a total of $L_C \times L_R^k=4$ number of non-zero elements in the cascaded VAD channel. Out of these $4$ non-zero elements, $L_C \times L_R=2$ is common across all the UEs due to the shared scatterer between the UEs and RIS. The partially-common row support for the first column is $\Omega_{r}^{1} = \{3\}$ and for the fourth column is $\Omega_{r}^{4}=\{1\}$. The common column sparsity along with the partially-common row sparsity is termed as the \textit{doubly-structured} sparsity.
\subsubsection{Individual sparsity} The scatterers which are not common across the different UEs, result in $\{\tilde{\mathbf{H}}_k\}_{k=1}^K$ having some UE-specific non-zero elements. In each non-zero column, there are $L_R^k-L_R$ non-zero elements which are not common across the different UEs corresponding to the UE-specific scatterers. This sparsity is termed as the individual sparsity of each user. In Fig. \ref{fig:double_sparsity}, scatterer 1 is specific to UE1 and results in one non-zero element in each non-zero column of $\tilde{\mathbf{H}}^1$ which is not common with $\tilde{\mathbf{H}}^2$. These are denoted by the blue boxes. The individual sparsity of UE2 due to scatterer 3 is denoted by the orange boxes.

Given the above sparsity of the cascaded VAD channel $\tilde{\mathbf{H}}$,
the problem of estimating $\tilde{\mathbf{H}}$, from the observation matrices
$\tilde{\mathbf{Y}}$, in \eqref{eq:system_model_stacked} now becomes a compressed sensing recovery problem. The authors in \cite{ris,amp} jointly estimated the cascaded VAD channel by compressed sensing recovery. The design in \cite{ris} jointly estimates the row and column supports of $\{\tilde{\mathbf{H}}^k\}_{k=1}^K$ and uses the OMP algorithm to get the channel estimates. The authors in \cite{amp} use the sparse Bayesian learning (SBL) framework which assumes a prior over the channel to encode its sparsity information. It assumed a Bernoulli-Gaussian family of prior on $\tilde{\mathbf{H}}$ and used VAMP algorithm for channel estimation. The design in \cite{ris}, however, is based on the OMP algorithm which requires the number of non-zero elements in $\tilde{\mathbf{H}}$ beforehand. The prior in \cite{amp} does not exploit the doubly-stuctured sparsity of the cascaded VAD channel exhaustively. It also suffers from time complexity issues as is the case with many SBL inference methods \cite{fmf-sbl}. In order to address these issues we propose two algorithms based on SBL inference method called VI for joint estimation of sparse channel in RIS-assisted mmWave systems. Unlike \cite{ris}, the proposed method does not require information about the number of non-zero elements in the channel and it also addresses the time complexity issues of \cite{amp}. We  show  that  the  proposed  designs based on VI and a column-wise coupled prior to encode the channel information,  due  to  better exploitation of channel sparsity, outperforms the one in \cite{ris,amp}.



\section{{Review of sparse Bayesian learning}} 
For the system model in \eqref{eq:system_model_stacked}, the Bayesian learning framework  treats $\tilde{\mathbf{Y}}$ and $ \tilde{\mathbf{H}}$ as the {observation} and unknown variables,  respectively.  The framework also assigns a \textit{prior} distribution $p(\tilde{\mathbf{H}}|\boldsymbol\Omega)$ to $\tilde{\mathbf{H}}$, with the hyperparameter set $\boldsymbol\Omega$.  The prior incorporates a belief about $\tilde{\mathbf{H}}$, which in this case is its sparsity structure. The Bayesian framework uses the likelihood distribution $p(\tilde{\mathbf{Y}}|\tilde{\mathbf{H}})$, along with the prior, to infer the posterior distribution of the unknown variable $\tilde{\mathbf{H}}$ using Bayes rule as {$p(\tilde{\mathbf{H}}|\tilde{\mathbf{Y}}, \boldsymbol\Omega) \propto p(\tilde{\mathbf{Y}} | \tilde{\mathbf{H}}) p(\tilde{\mathbf{H}}|\boldsymbol\Omega)$}. The choice of prior distribution is critical in Bayesian learning framework. Recent literature in {\cite{sbl,msbl,pcsbl,Ke2020}} has considered different kind of sparsity-promoting priors. We now discuss these works, and the limitations of priors used therein in capturing the doubly-structured sparsity in the cascaded VAD channel.

\subsubsection{{SBL}} The authors in \cite{sbl} recovered  a sparse vector $\tilde{\mathbf{h}}_m$ from the observation $\tilde{\mathbf{y}}_m = \tilde{\boldsymbol{\Theta}} \tilde{\mathbf{h}}_m +\tilde{\mathbf{n}}$, a problem which can be obtained by setting $M = 1$ in \eqref{eq:system_model_stacked}. The design in \cite{sbl} assigned a zero-mean Gaussian prior to the sparse vector $ \tilde{\mathbf{h}}_m $ i.e, $p( \tilde{\mathbf{h}}_m |\boldsymbol{\alpha}) = \prod_{n=1}^{NK} \mathcal{CN} (\tilde{h}_{nm} |0, \alpha_n^{-1}),$
with $\boldsymbol{\alpha} =[\alpha_1,\hdots , \alpha_{NK}]^T \in \mathbb{R}_+^{NK\times 1}$ 
being the {precision} (inverse of variance) hyperparameter vector. When $\alpha_n$ is  high, the unknown variable {$\tilde{h}_{nm}$, with a very high probability,  takes a value close to its mean, which is zero. The zero-mean Gaussian prior, with independent entries, thus only captures the individual sparsity of each entry, but not the common column and partially common-row sparsity of the cascaded VAD channel.

	\subsubsection{\textit{cCP-SBL}} The authors in \cite{cp-sbl} considered a centralized coupled prior algorithm for joint active UE detection and channel estimation.
	This work assumed a generalized prior for the composite channel $\tilde{\mathbf{H}}$. The prior on its $(n,m)$th element is given as
	\begin{align}
		\label{eq:prior2a}
		p(\tilde{h}_{nm}\vert \boldsymbol{\alpha}_n, \boldsymbol{\beta}_{nm})&=\mathcal{CN}\Big(\tilde{h}_{nm}\Big\vert 0,\left(\sum_{b=1}^{M} \beta_{n,m,b}\alpha_{n,b}\right)^{-1} \Big)=\frac{ \boldsymbol{\beta}_{nm}^T {                                                                                                                                                                                                                                                                                                                                                                                                                                                                                                                                                                                                                                                                                                                                                                                                                                                                                                         \boldsymbol{\alpha}_n}}{\pi}\exp\frac{-|\tilde{h}_{nm}|^2}{(\boldsymbol{\beta}_{nm}^T \boldsymbol{\alpha}_n)^{-1}}.
	\end{align}
	
	The non-negative scalar $\alpha_{nm}$ is the precision hyperparameter for $\tilde{h}_{nm}$. The scalar $\beta_{n,m,b}$ is the shared weight hyperparameter, which denotes the relevance given to  $\alpha_{n,b}$ to calculate the variance of  $\tilde{h}_{nm}$.  The column vector  $\boldsymbol{\alpha}_n=[\alpha_{n,1}, \alpha_{n,2}, \cdots, \alpha_{nm} ]^T \in \mathbb{R}_+^{ M \times 1}$ and the row vector $\boldsymbol{\beta}_{nm}^T=[ \beta_{n,m,1} , \beta_{n,m,2} , \cdots , \beta_{n,m,M} ] \in[0,1]^{1\times M}$ thus denote the precision and shared weight vector for the $n$th row, respectively. 
		The shared weight parameters $\beta_{n,m,b}$ of \eqref{eq:prior2a}, however, are such that it wrongly models the shared sparsity among the elements of a row in $\tilde{\mathbf{H}}$, and ignores any shared sparsity across the rows. It, thus, captures the  individual sparsity for each UE but fails to capture the common column and partially-common row sparsity of $\tilde{\mathbf{H}}$, which arises due to the shared sparsity across different rows (UEs).

		\subsubsection{\textit{FMF-SBL}} Worley in \cite{fmf-sbl}  proposed a faster verison of the SBL algorithm which is scalable in the number of measurements. For the system $\tilde{\mathbf{y}}_m = \tilde{\boldsymbol{\Theta}} \tilde{\mathbf{h}}_m +\tilde{\mathbf{n}}$, the algorithm finds the posterior distribution over $\tilde{\mathbf{h}}_m$, i.e. $p(\tilde{\mathbf{h}}_m | \tilde{\mathbf{y}})$ by approximating it as $    p(\tilde{\mathbf{h}}_m | \tilde{\mathbf{y}}) \approx q(\tilde{\mathbf{h}}_m)$, with 
		$q(\tilde{\mathbf{h}}_m)$ being the approximating posterior distribution. The design in \cite{fmf-sbl} optimizes the parameters of $q(\tilde{\mathbf{h}}_m)$ to make it close to the true posterior distribution. This design further simplify the posterior calculation  by assuming a fully-factorized approximating posterior distribution as follows:
		\begin{align}
			q(\tilde{\mathbf{h}}_m) = \prod_{n=1}^{NK} q(\tilde{h}_{nm}).
		\end{align}
		Using the above approximations,  this work proposed the fast mean-field (FMF)-SBL algorithm. This proposed algorithm, however, assumed that the prior over $\tilde{\mathbf{h}}_m$ is a simple multivariate Gaussian with a diagonal covariance matrix. This prior, thus, only captures the individual sparsity.

		\section{Proposed column-wise coupled prior}
		We now propose a column-wise coupled prior which captures the complete sparsity of the cascaded VAD channels of all UEs i.e. $\{\tilde{\mathbf{H}}^{k}\}_{k=1}^{K}$. The  sparsity can be summarized as follows:
		\begin{itemize}
			\item the channel matrix $\tilde{\Hmat}^k$ has $L_C$ non-zero columns, and the column support is common across all $\tilde{\Hmat}^k,\forall k = 1, \hdots, K$ UEs.
			\item In each non-zero column of $\{\tilde{\mathbf{H}}^{k}\}_{k=1}^{K}$, there are $L_R^k$ non-zero rows, out of which $L_R$, with $L_R \leq L_R^k$, non-zero row entries in each column being common to all the UEs. There are, thus, $L_R^k-L_R$ non-zero row entries which are UE-specific. 
		\end{itemize}
		
		We assume a column-wise coupled prior over the joint cascaded VAD channel $\tilde{\Hmat}$. This prior, as we will discuss later, models the doubly-structured and individual sparsity of $\tilde{\Hmat}$. Each element of the channel $\tilde{\mathbf{H}}$, denoted as $\tilde{h}_{nm}$, is assumed to be generated from the following prior:
		\begin{align}
			\label{eq:prior}
			p(\tilde{h}_{nm}\vert \boldsymbol{\alpha}_m, \mathbf{u}_{m})
			&=\mathcal{CN}\left(\tilde{h}_{nm}\Bigg\vert 0,\left(u_{nm} \alpha_{m}^{\text{NZ}}+{(1-u_{nm})} \alpha_{m}^{\text{Z}}\right)^{-1} \right) \nonumber \\ 
			&=\frac{1}{\pi\gamma_{nm}}\exp\frac{-|\tilde{h}_{nm}|^2}{\gamma_{nm}}.
		\end{align}
		The scalar $u_{nm}$ denotes the support of $\tilde{h}_{nm}$ which satisfies $u_{nm} = \mathbb{I}(\tilde{h}_{nm} \neq 0)$.	The non-negative scalars $\alpha_{m}^{\text{NZ}}$ and $\alpha_{m}^{\text{Z}}$ represent the precision parameter of $\tilde{h}_{nm}$, when it is non-zero and zero, respectively. 
		The scalar $\gamma_{nm}$ is the variance of $\tilde{h}_{nm}$, which is defined as follows:
		\begin{align}\label{eq:prior_variance}
			\gamma_{nm} \triangleq  \left({u_{nm}}\alpha_{m}^{\text{NZ}}+{(1-u_{nm})} \alpha_{m}^{\text{Z}}\right)^{-1}.
		\end{align}
		The column vector $\boldsymbol{\alpha}_{m} = [\alpha_{m}^{\text{NZ}},\alpha_{m}^{\text{Z}}]^{T}\in \mathbb{R}_{+}^{2 \times 1}$ denotes the prior precision parameter for the $m$th column of $\tilde{\mathbf{H}}$. The column vector $\mathbf{u}_m=[u_{1,m},u_{2,m},\dots,u_{NK,m}]^{T}\in \{0,1\}^{NK \times 1}$ is the support vector for the $m$th column of $\tilde{\mathbf{H}}$. 
		The prior distribution for the joint cascaded VAD channel is thus given as,
		\begin{align}
			\label{eq:prior2}
			p(\tilde{\mathbf{H}}\vert \mathbf{A},\mathbf{U} ) 
			= \prod_{n=1}^{NK} \prod_{m=1}^{M} p(\tilde{h}_{nm}\vert \boldsymbol{\alpha}_m,\mathbf{u}_{m})
			= \prod_{m=1}^{M} \mathcal{CN}(\tilde{\mathbf{h}}_{m}\vert \mathbf{0}_{NK}, \boldsymbol{\Upsilon}_m  ).
		\end{align}
		The matrix $\mathbf{A} =[\boldsymbol{\alpha}_1, \hdots, \boldsymbol{\alpha}_M]\in\mathbb{R}_+^{2\times M}$ is the precision parameter matrix, and $\mathbf{U}=[\mathbf{u}_1,\hdots,\mathbf{u}_m] \in \{0,1\}^{NK \times M}$ is the binary support matrix. The diagonal matrix $\boldsymbol{\Upsilon}_m= \text{diag}(\boldsymbol{\gamma}_m )$ denotes the covariance matrix of the $m$th column of $\tilde{\mathbf{H}}$, i.e. $\tilde{\mathbf{h}}_{m}$. The $n$th entry of the vector $\boldsymbol{\gamma}_m$,  i.e. $\gamma_{nm}$, is  the variance of $\tilde{h}_{nm}$, and is defined in \eqref{eq:prior_variance}.

		

		\subsection{Discussion on how prior parameters capture the doubly-structured sparsity}
		\label{intuition}
		We note from \eqref{eq:prior} that 
		\begin{itemize}
			\item 	all the non-zero entries $\tilde{h}_{nm}\neq 0, \forall n = 1, \hdots, NK$ in the $m$th column of $\tilde{\Hmat}$ (i.e.,  all $n$ such that $u_{nm} = 1$)  are drawn from the a Gaussian distribution  $\CNcal(\mathbf{0}_M, \alpha_m^{\text{NZ}})$. 
			
			\item  all the zero entries $\tilde{h}_{nm}= 0, \forall n = 1, \hdots, NK$ in the $m$th column of $\tilde{\Hmat}$ (i.e.,  all $n$ such that $u_{nm} = 0$)  are drawn from the a Gaussian distribution  $\CNcal(\mathbf{0}_M, \alpha_m^{\text{Z}})$. 
		\end{itemize} 
		The prior precision hyperparameters  $\alpha_{m}^{\text{NZ}}$ and $\alpha_{m}^{\text{Z}}$ are same for all $n = 1, \hdots, NK$ rows.  This ensures that the prior distribution of the entries $\tilde{h}_{nm}$ for all $n = 1, \hdots, NK$, which share same sparsity structure are drawn from same prior distribution. We couple  all the non-zero (zero) entries in the column $\tilde{\mathbf{h}}_m$ by assuming same prior distribution over them. The non-zero and zero entries are found using the parameter $u_{nm}$, the support of $\tilde{h}_{nm}$. Also, since the hyperparameter $\alpha_m^{\text{NZ}}$ corresponds to the case when $u_{nm} = 1$ or $\tilde{h}_{nm} \neq 0$, it has a value close to zero. Similarly, the hyperparameter $\alpha_m^{\text{Z}}$. The prior proposed in \eqref{eq:prior2} captures the 
		\begin{enumerate}
			\item \textit{common-column sparsity} by having a shared hyperparameters $\alpha_m^{\text{NZ}}$, $\alpha_m^{\text{Z}}$, which are independent of the entries in the column. When the column $m$ is completely zero, $\tilde{\hv}_m = \mathbf{0}$, $u_{nm} = 0, \forall n$. The entries $\tilde{h}_{nm}, \forall n $,  from \eqref{eq:prior}, are thus  drawn from  $\CNcal(\mathbf{0}_M, \alpha_m^{\text{Z}})$, where $\alpha_m^{\text{Z}} \rightarrow \infty$, which makes all $\tilde{h}_{nm} \rightarrow 0,$ for all $n = 1, \hdots, NK$ entries in the $m$th column.
			
			\item \textit{partially-common row} and \textit{individual row sparsity} by assigning same distribution $\CNcal(\mathbf{0}_M, \alpha_m^{\text{NZ}})$ to all the entries in the $m$th column which are non-zero; and same distribution $\CNcal(\mathbf{0}_M, \alpha_m^{\text{Z}})$ to all the entries in the $m$th column which are equal to zero.
		\end{enumerate}
		The proposed prior does not distinguish between the partially-common sparsity across different UEs and the individual row sparsities. It instead captures the shared sparsity in the complete column $\tilde{\mathbf{h}}_m =[\tilde{h}_{1,m}, \hdots, \tilde{h}_{NK,m}]^T$, which includes both the sparsities. The prior is thus generalized, which is able to capture any kind of shared sparsity in the non-zero columns of the channel.

		The parameters of the proposed prior, i.e., the precision matrix $\mathbf{A}$ and the binary support matrix $\mathbf{U} = [\mathbf{u}_1,\hdots,\mathbf{u}_M] \in \{0,1\}^{NK \times M}$, need to be estimated. To facilitate estimation of $\mathbf{A}$, similar to \cite{tipping}, we assign a Gamma \textit{hyperprior} distribution over its elements:
		\begin{align}
			p(\mathbf{A})&=\prod_{i\in\{NZ,Z\}}\prod_{m=1}^{M} p(\alpha_{m}^i) = \prod_{i\in\{NZ,Z\}} \prod_{m=1}^{M} \Gamma(c)^{-1} d^{c}(\alpha_{m}^i)^{ c } e^{- d  \alpha_{m}^{i}} .\label{eq:A_prior}
		\end{align}
		Here, $\Gamma(c)=\int_0^{\infty}t^{c-1}e^{-t}dt$ is the Gamma function.  The Gamma distribution is a suitable hyperprior for the positive quantity, precision, and makes the Bayesian inference tractable \cite{tipping}.  To promote sparsity in the entries of $\tilde{\mathbf{H}}$ we choose the value of $c=1$ and $d=10^{-8}$ \cite{cp-sbl}, as they encourage large values of precision parameters $\alpha_{nm}$, and consequently close-to-zero values of
		channel entries $\tilde{h}_{nm}$. With $p(\tilde{\mathbf{H}}|\mathbf{A}, \mathbf{U})$ from \eqref{eq:prior2}, and $p(\mathbf{A})$ from \eqref{eq:A_prior}, the prior $ p(\tilde{\mathbf{H}}| \mathbf{U})$ on $\tilde{\mathbf{H}}$, which is calculated as $ p(\tilde{\mathbf{H}}| \mathbf{U}) =  \int p(\tilde{\mathbf{H}}|\mathbf{A}, \mathbf{U}) p(\mathbf{A}) d\mathbf{A}$, has a Student-t distribution {\cite{tipping}}.
		This distribution, due to its sharp peak at zero \cite{tipping},  has a better sparsity-encouraging properties than the Gaussian prior.   {This approach for constructing a prior using hyperprior is called hierarchical prior modeling, which allows us to indirectly use complex non-conjugate prior distributions  for tractable inference \cite{bishop}.} In order to estimate $\mathbf{U}$ we use the log-likelihood ratio test similar to \cite{cp-sbl}. The updates for $\mathbf{A}$ and $\mathbf{U}$ are dependent on one another, hence, we use an alternating optimization technique to estimate them
			along with the channel $\tilde{\mathbf{H}}$. We  propose an iterative variational expectation maximization (VEM) approach for their estiamtion \cite{vem}. The VEM approach incorporates the idea of variational inference in the EM algorithm. 

		\vspace{-0.21cm}
		\section{Variational expectation maximization approach with coupled prior}\label{prior}
	VI is a method to infer the posterior distribution over all the unknowns in the model. It helps in dealing with intractable posteriors and casts the inference problem as an optimization problem. The idea is to posit a simple family of distributions over the unknowns and find the member of the family that is closest in KL divergence to the true posterior distribution. It treats $\tilde{\mathbf{Y}}$ and $ \tilde{\mathbf{H}}$ as the {observation} and unknown variables,  respectively. The parameters of the model are prior precision matrix $\mathbf{A}$ and the binary support matrix $\mathbf{U}$. VI infers the posterior distribution over all the unknowns {$p(\tilde{\mathbf{H}},\mathbf{A},\mathbf{U}|\tilde{\mathbf{Y}})$}. This inference problem is, however, difficult to solve due to the complex structure of the posterior distribution \cite{fmf-sbl}. In order to simplify this inference problem variational expectation maximization (VEM) is used which infers the posterior distribution over some unknowns and finds point estimates for the rest. We use the VEM framework to estimate the posterior distribution over $\tilde{\mathbf{H}}$ denoted as $p(\tilde{\mathbf{H}}|\tilde{\mathbf{Y}},\mathbf{A},\mathbf{U})$ and find point estimates for $\mathbf{A},\mathbf{U}$. Since $\mathbf{A}\text{ and }\mathbf{U}$ are parameters common to the entire model, their uncertainty estimates are quite low and hence doing just a point estimation over them suffices \cite{vem}.
		The VEM-algorithm is a two-step iterative procedure based on the following identity \cite{vem}:
				\begin{align}\label{eq:variational_identity}
						p(\tilde{\mathbf{Y}}\vert\mathbf{A},\mathbf{U}) &= \mathcal{L}(q,\mathbf{A},\mathbf{U}) + \text{KL}(q || p), \;\text{where,} \\
						\mathcal{L}(q,\mathbf{A},\mathbf{U}) &= \int q(\tilde{\mathbf{H}}) \ln(\frac{p(\tilde{\mathbf{Y}},\tilde{\mathbf{H}}\vert\mathbf{A},\mathbf{U})}{q(\tilde{\mathbf{H}})}) d\tilde{\mathbf{H}}, 
						\label{eq:ELBO_definition}
						\text{KL}(q || p) &= - \int q(\tilde{\mathbf{H}}) \ln \frac{p(\tilde{\mathbf{H}}|\tilde{\mathbf{Y}},\mathbf{A},\mathbf{U})}{q(\tilde{\mathbf{H}})}) d\tilde{\mathbf{H}}.
					\end{align}
	    Here KL$(q||p)$ is the KL-divergence between $q$ and $p$, $q$ is any probability distribution over $\mathbf{\tilde{H}}$ and $\mathcal{L}(q,\mathbf{A},\mathbf{U})$ is called the ELBO. VEM solves an optimization over a class of tractable distributions $Q$ in order to find an approximating distribution $q \in Q$, that is most similar to $p$. Let us denote the parameter set of the model as $\boldsymbol{\Theta} =\{\mathbf{A},\mathbf{U}\}$. Let us denote the approximating distribution for $p(\tilde{\mathbf{H}}|\tilde{\mathbf{Y}},\mathbf{A},\mathbf{U})$ as $q(\tilde{\mathbf{H}};\boldsymbol{\theta})$. The E-step of VEM algorithm, calculates $q(\tilde{\mathbf{H}};\boldsymbol{\theta})$  such that it maximizes the ELBO  $\mathcal{L}(q,\boldsymbol{\Theta})$ for a fixed $\boldsymbol{\Theta}$. To perform this maximization, a particular form of $q(\tilde{\mathbf{H}};\boldsymbol{\theta})$ must be assumed, so that the problem of inference of $q(\tilde{\mathbf{H}};\boldsymbol{\theta})$ can be reduced to the optimization of its parameters $\boldsymbol{\theta}$ \cite{vem}. The parameters, $\boldsymbol{\theta}$, of the approximating posterior distribtuion are called the variational parameters. The ELBO $\mathcal{L}(\boldsymbol{\theta},\boldsymbol{\Theta})$, thus, becomes a function of these parameters and is maximized with respect to $\boldsymbol{\theta}$
	    in the E-step, and with respect to $\boldsymbol{\Theta}$ in the M-step \cite{vem}. The M-step of the VEM algorithm is shown to be equivalent to maximizing the expected complete data log-likelihood (CLL) of the model defined as $\langle p(\tilde{\mathbf{H}},\tilde{\mathbf{Y}};\boldsymbol{\Theta}) \rangle_{q(\tilde{\mathbf{H}};\boldsymbol{\theta}^{new})}$ \cite{vem}. The VEM algorithm is summarized below,
	    \begin{align}\label{eq:expected_CLL}
	    	&\text{E-step: Evaluate }\boldsymbol{\theta}^{new}=\arg \max_{\boldsymbol{\theta}} \mathcal{L}(q(\tilde{\mathbf{H}};\boldsymbol{\theta}),\boldsymbol{\Theta}^{old})\\ \nonumber
	    	&\text{M-step: Evaluate } \boldsymbol{\Theta}^{new} = \arg \max_{\boldsymbol{\Theta}} \langle p(\tilde{\mathbf{H}},\tilde{\mathbf{Y}};\boldsymbol{\Theta}) \rangle_{q(\tilde{\mathbf{H}};\boldsymbol{\theta}^{new})}.
	    \end{align}
	    Employing this VEM algorithm for channel estimation in RIS-aided mmWave systems not only allows us to deal with intractable posterior distributions but also, as seen later,  allows us to obtain faster updates for $\boldsymbol{\theta}$ by assuming simpler forms of the approximating distribution $q(\tilde{\mathbf{H}};\boldsymbol{\theta})$.
	    In order to obtain updates for $q(\tilde{\mathbf{H}};\boldsymbol{\theta})$, we derive the ELBO expression which is optimized to get  variational parameters $\boldsymbol{\theta}$. Based on the structure of $q(\tilde{\mathbf{H}};\boldsymbol{\theta})$ we propose two algorithms, the SM-SBL and FM-SBL algorithm.  The SM-SBL algorithm assumes a structured form of the approximating distribution and optimizes the ELBO to obtain updates for the structured variational parameters. The FM-SBL algorithm assumes a factorized approximating distribution and optimizes a bound on the ELBO to obtain faster iterative updates for factorized variational parameters. It is shown in Appendix \ref{app:convergence} that the proposed updates ascend on the actual ELBO. 

		\vspace{-0.3cm}
		\subsection{ELBO expression with the proposed coupled prior}
		We begin by calculating the ELBO expression with the coupled prior of \eqref{eq:prior2}. This ELBO expression is optimized in the E-step of the VEM algorithm to get the variational parameters $\boldsymbol{\theta}$. As discussed earlier we will need to assume an appropriate form for the approximating distribution of the posterior, so that the ELBO can be maximized with respect to its parameters. 
		In order to obtain the ELBO we need the likelihood distirbution $p(\tilde{\mathbf{Y}}\vert \tilde{\mathbf{H}})$. It is defined as follows \cite{cp-sbl}:
		\begin{align}
			\label{eq:ris_likelihood}
			p(\tilde{\mathbf{Y}}\vert \tilde{\mathbf{H}}) & = \prod_{m=1}^{M} p(\tilde{\mathbf{y}}_{m}\vert \tilde{\mathbf{h}}_{m}) = \prod_{m =1 }^M \mathcal{CN} (\tilde{\mathbf{y}}_{m}\vert  \tilde{\boldsymbol{\Theta}} \tilde{\mathbf{h}}_{m}, \sigma^2\mathbf{I}_{Q}).
		\end{align}
		The posterior distribution of $\tilde{\mathbf{H}}$ which needs to be approximated has the following form \cite{cp-sbl}:
		\begin{align}
			p(\tilde{\mathbf{H}} \vert \tilde{\mathbf{Y}},\mathbf{A},\mathbf{U}) & = \prod_{m = 1}^M p(\tilde{\mathbf{h}}_{m}|\tilde{\mathbf{y}}_{m}, \mathbf{A}, \mathbf{U}).
		\end{align}
		To estimate $p(\tilde{\mathbf{h}}_{m}\vert \tilde{\mathbf{y}}_{m},\mathbf{A},\mathbf{U})$, we employ the VEM technique discussed earlier. To this end, the approximating posterior distribution for $p(\tilde{\mathbf{h}}_{m}\vert \tilde{\mathbf{y}}_{m},\mathbf{A},\mathbf{U})$ is assumed as $q(\tilde{\mathbf{h}}_m;\boldsymbol{\theta})$. Here $\boldsymbol{\theta}$ denotes the variational parameters, whose expressions are calculated by maximizing the ELBO. In our model the parameters are $\boldsymbol{\Theta}=\{\mathbf{A},\mathbf{U}\}$. The ELBO expression from \eqref{eq:ELBO_definition} for the model defined can be written as follows :
		\begin{align}\label{eq:common_ELBO}
			\mathcal{L}[q(\tilde{\mathbf{h}}_m;\boldsymbol{\theta}),\mathbf{A},\mathbf{U}] &= \mathbb{E}_{q}[\ln p(\tilde{\mathbf{y}}_m,\tilde{\mathbf{h}}_m)] - \mathbb{E}_q[\ln q(\tilde{\mathbf{h}}_{m};\boldsymbol{\theta})] \\ \nonumber
			&\stackrel{(a)}{=}\mathbb{E}_{q}[\ln p(\tilde{\mathbf{y}}_m \vert \tilde{\mathbf{h}}_m)] + \mathbb{E}_{q}[\ln p(\tilde{\mathbf{h}}_m)] - \mathbb{E}_q[\ln q(\tilde{\mathbf{h}}_{m};\boldsymbol{\theta})] \\ \nonumber
			&\stackrel{(b)}{\propto} -\frac{1}{\sigma^2}\langle||\tilde{\mathbf{y}}_m - \tilde{\boldsymbol{\Theta}}\tilde{\mathbf{h}}_m||_2^{2}\rangle-\sum_{n=1}^{NK}\langle|\tilde{h}_{nm}|^{2}\rangle(\gamma_{nm})^{-1} - \langle \ln q(\tilde{\mathbf{h}}_{m};\boldsymbol{\theta})\rangle. \\ \nonumber
			&= -\frac{1}{\sigma^2}\lambda(\boldsymbol{\theta})-\sum_{n=1}^{NK}\beta_n(\boldsymbol{\theta})(\gamma_{nm})^{-1} + H(\boldsymbol{\theta}).
		\end{align}
		Here 
		$	\lambda(\boldsymbol{\theta}) = \langle||\tilde{\mathbf{y}}_m - \tilde{\boldsymbol{\Theta}}\tilde{\mathbf{h}}_m||_2^{2}\rangle, \;
		\beta_n(\boldsymbol{\theta}) = \langle|\tilde{h}_{nm}|^{2}\rangle \text{, and } 
		H(\boldsymbol{\theta}) = - \langle \ln q(\tilde{\mathbf{h}}_{m};\boldsymbol{\theta})\rangle$. The expectations are with respect to the approximating distribution $q(\tilde{\mathbf{h}}_m;\boldsymbol{\theta})$.
		Equality in $(a)$ is obtained by replacing $p(\tilde{\mathbf{y}}_m,\tilde{\mathbf{h}}_m)$ with $p(\tilde{\mathbf{y}}_m \vert \tilde{\mathbf{h}}_m)p(\tilde{\mathbf{h}}_m)$. Proportionality in (b) is obtained by substituting the likelihood and prior from \eqref{eq:ris_likelihood} and \eqref{eq:prior} respectively, and by considering only the terms dependent on the variational parameters.
		
		Based on the form of the approximating distribution $q(\tilde{\mathbf{h}}_{m};\boldsymbol{\theta})$ , the ELBO expression in \eqref{eq:common_ELBO} take different forms. We next propose structured mean field sparse Bayesian learning algorithm (SM-SBL) and factorized mean field sparse Bayesian learning algorithm (FM-SBL) algorithms which exploit the VEM technique. The SM-SBL algorithm assumes a structured version of the approximating distribution $q(\tilde{\mathbf{h}}_{m};\boldsymbol{\theta})$ and maximizes the ELBO to calculate the parameters of the structured $q(\tilde{\mathbf{h}}_{m};\boldsymbol{\theta})$. This form of structured distribution leads to a prohibitively expensive updates for the parameters of $q(\tilde{\mathbf{h}}_{m};\boldsymbol{\theta})$. To circumvent this issue, we propose a factorized version of $q(\tilde{\mathbf{h}}_{m};\boldsymbol{\theta})$ in the FM-SBL algorithm. This technique, in addition to assuming a factorized version of the approximating distribution $q(\tilde{\mathbf{h}}_{m};\boldsymbol{\theta})$, also uses the Lipschtiz inequality to lower bound the ELBO. It then maximzes this lower bound to estimate the parameters of the factorized $q(\tilde{\mathbf{h}}_{m};\boldsymbol{\theta})$. Optimizing the lower bound of the ELBO instead of the actual ELBO leads to faster updates of the variational parameters.
		
		\subsection{Structured mean field sparse Bayesian learning (SM-SBL) algorithm}
		The SM-SBL algorithm assumes the following structured approximate posterior distribution:
		\begin{align}\label{eq:approximating_posterior_form}
			q(\tilde{\mathbf{h}}_{m};\boldsymbol{\theta}) = {\mathcal{CN} }(\tilde{\mathbf{h}}_{m}\vert\boldsymbol{\mu}_m,\boldsymbol\Sigma_{m}).
		\end{align}
		The mean and covariance matrix $\boldsymbol{\mu}_m \text{ and }\boldsymbol{\Sigma}_m$ respectively, of the multivariate normal distribution $q$  are its variational parameters. The E-step of SM-SBL algorithm calculates the variational parameters by maximizing the ELBO in \eqref{eq:common_ELBO}. The M-step calculates the parameters $\boldsymbol{\Theta}$, i.e., the precision matrix $\mathbf{A}$ and the binary support matrix $\mathbf{U}$.
		
		\subsubsection{{E-step of the SM-SBL algorithm}}
		The E-step assumes that $\mathbf{A}$ and $\mathbf{U}$ are fixed at their current values and optimizes the ELBO with respect to $\boldsymbol{\mu}_m\text{ and }\boldsymbol\Sigma_{m}$. For the structured distribution in \eqref{eq:approximating_posterior_form}, the ELBO expression is calculated in the next lemma, which is derived in Appendix \ref{app:lambda_structured}. For notational simplicity we omit $\mathbf{A}$ and $\mathbf{U}$ from the ELBO on left hand side since they are fixed in the E-step.
		\begin{lemma}
			The ELBO expression for the structured distribution in  \eqref{eq:approximating_posterior_form} can be written as follows:
			\begin{align}\label{eq:ELBO_structured}
				&\mathcal{L}_S[\boldsymbol{\mu}_m,\boldsymbol\Sigma_{m}] = -\frac{1}{\sigma^2}\lambda(\boldsymbol{\mu}_m,\boldsymbol\Sigma_{m})-\sum_{n=1}^{NK}\beta_n(\boldsymbol{\mu}_m,\boldsymbol\Sigma_{m})(\gamma_{nm})^{-1} + H(\boldsymbol{\Sigma}_m)  \\ 
				&=-\frac{1}{\sigma^2}||\tilde{\mathbf{y}}_m - \tilde{\boldsymbol{\Theta}}\tilde{\boldsymbol{\mu}}_m||_2^{2}-\frac{1}{\sigma^2}\text{tr}(\tilde{\boldsymbol{\Theta}}^H\tilde{\boldsymbol{\Theta}}\boldsymbol{\Sigma}_m)-\sum_{n=1}^{NK}(\vert\mu_{nm}\vert^2+\tau_{nm})(\gamma_{nm})^{-1} + \ln \det(\boldsymbol{\Sigma}_m), \notag \end{align}
			where 
			$\lambda(\boldsymbol{\mu}_m,\boldsymbol\Sigma_{m})  = ||\tilde{\mathbf{y}}_m - \tilde{\boldsymbol{\Theta}}\tilde{\boldsymbol{\mu}}_m||_2^{2} +  \text{tr}(\tilde{\boldsymbol{\Theta}}^H\tilde{\boldsymbol{\Theta}}\boldsymbol{\Sigma}_m), 
			\beta_n(\boldsymbol{\mu}_m,\boldsymbol\Sigma_{m}) = \vert\mu_{nm}\vert^2+\tau_{nm}, \text{ and }
			H(\boldsymbol{\Sigma}_m) = \ln \det(\boldsymbol{\Sigma}_m)$.
		\end{lemma}
		To calculate the update expression for the variational parameters $\boldsymbol{\mu}_m$ and $\boldsymbol{\Sigma}_m$, the ELBO is next maximized using the first order stationarity conditions \cite{fmf-sbl}.  We, therefore,  calculate the gradients of ELBO with respect to $\boldsymbol{\mu}_m$ and $\boldsymbol{\Sigma}_m$, which are given as follows:
		\begin{align}\label{eq:ELBO_gradients_CPSBL}
			\nabla_{\boldsymbol{\mu}_m} \mathcal{L}_S[\boldsymbol{\mu}_m,\boldsymbol\Sigma_{m}] &= -\frac{2}{\sigma^2}(\tilde{\boldsymbol{\Theta}}^H\tilde{\boldsymbol{\Theta}}\boldsymbol{\mu}_m-\tilde{\boldsymbol{\Theta}}^H\tilde{\mathbf{y}}_m)-2\; \text{diag}(\boldsymbol{\gamma}_m)^{-1},\\ \nonumber
			\nabla_{\boldsymbol{\Sigma}_m} \mathcal{L}_S[\boldsymbol{\mu}_m,\boldsymbol\Sigma_{m}] &= -\frac{1}{\sigma^2}\tilde{\boldsymbol{\Theta}}^H\tilde{\boldsymbol{\Theta}}- \text{diag}(\boldsymbol{\gamma}_m)^{-1} +\boldsymbol{\Sigma}_m^{-1}.
		\end{align}
		
		The update expression can now be calculated by equating these gradients to zero.
		\begin{align}
			\label{eq:W_cov}
			\boldsymbol\Sigma_{m}& =\left( \sigma^{-2} \tilde{\boldsymbol{\Theta}}^{H}\tilde{\boldsymbol{\Theta}}+ \text{diag}(\boldsymbol{\gamma}_m)^{-1} \right)^{-1} \;\text{ and }
			\boldsymbol{\mu}_m  = \sigma^{-2} \boldsymbol\Sigma_{m}\tilde{\boldsymbol{\Theta}}^{H}\tilde{\mathbf{y}}_{m}.
		\end{align}
		

		
		\subsubsection{{M-step of the SM-SBL algorithm}} The M-step of SM-SBL algorithm estimates the precision matrix $\textbf{A}$ and the support matrix $\textbf{U}$ by employing alternating maximization. This is because they depend on each other. The matrix $\textbf{A}$ is calculated by assuming $\textbf{U}$ to be a constant, and vice-versa.
		
		$\bullet $ \textbf{Estimation of  precision matrix $\mathbf{A}$:} We saw that in the M-step of VEM algorithm the parameters are estimated by maximizing the expected CLL of \eqref{eq:expected_CLL}. The prior over the precison matrix $\mathbf{A}$ is given by \eqref{eq:A_prior}, thus, CLL for our model is defined as $[\log p(\tilde{\mathbf{Y}},\tilde{\mathbf{H}}|\mathbf{A},\mathbf{U}) + \log p(\mathbf{A})]=\log p(\mathbf{A},\tilde{\mathbf{Y}},\tilde{\mathbf{H}}|\mathbf{U})$. Due to the coupling in the parameters of $\mathbf{A}$, the optimization problem for expected CLL does not have a closed form solution. We try to obtain a sub-optimal range of solutions by analyzing the optimiality conditions. In order to estimate $\mathbf{A}$ we fix the value of $\mathbf{U}$ and maximize the expected CLL of \eqref{eq:expected_CLL} as follows:
		\begin{align}
			\hat{\mathbf{A}} {=} \arg \underset{\mathbf{A}}{\max} \, \mathbb{E}_{\tilde{\mathbf{H}}|\tilde{\mathbf{Y}},\mathbf{A}^{old},\mathbf{U}}[\log p(\mathbf{A},\tilde{\mathbf{Y}},\tilde{\mathbf{H}}|\mathbf{U})].
		\end{align}
		The expectation is taken with respect to the posterior distribution calculated in the E-step. In the above expression, $\mathbf{A}^{old}$ represents its current estimate. The above maximization problem can equivalently be cast as follows:
		\begin{align} 
			&\max_{\mathbf{A}}    \mathbb{E}_{  \tilde{\mathbf{H}}} \left[\log \left( p(\mathbf{A}, \tilde{\mathbf{Y}}, \tilde{\mathbf{H}} |\mathbf{U} \right)\right] \nonumber 
			{=}  \max_{\mathbf{A}}    \mathbb{E}_{  \tilde{\mathbf{H}}}  \left[\log \left( p(\mathbf{A}) p(\tilde{\mathbf{Y}}\vert \tilde{\mathbf{H}})p(\tilde{\mathbf{H}} | \mathbf{A},\mathbf{U}) \right)\right]\nonumber\\
			& \stackrel{(a)}{=} \max_{\mathbf{A}} \sum_{i\in \{NZ,Z\}}\sum_{m=1}^{M}\Big(c \log \alpha_{m}^i - d \alpha_{m}^i\Big)+\Big(\sum_{n=1}^{NK}\sum_{m=1}^{M}\log (\gamma_{nm} )^{-1}-(\gamma_{nm} )^{-1} \mathbb{E}_{  \tilde{\mathbf{H}} } \Big[|\tilde{h}_{nm}|^2\Big]\Big).\label{eq:max_surrogate}
		\end{align}
		For notational simplicity, we use the notation $\mathbb{E}_{\tilde{\mathbf{H}}} [\cdot] = \mathbb{E}_{\tilde{\mathbf{H}}|\tilde{\mathbf{Y}}, \mathbf{A}^{old}} [\cdot] $.  Equality $(a)$ is obtained by i) ignoring the distribution $p(\tilde{\mathbf{Y}}|\tilde{\mathbf{H}})$ as it is independent of $\mathbf{A}$; and ii) using \eqref{eq:A_prior}  and \eqref{eq:ris_likelihood}. We note that the optimization in \eqref{eq:max_surrogate} cannot be decoupled into $2\times M$ problems,  for each hyperparameter $\alpha_{m}^i$. This is because $\alpha_{m}^i, \forall m $, are coupled together in the logarithm term $\log (\gamma_{nm} )^{-1}$. For such cases, a closed form solution of \eqref{eq:max_surrogate} is difficult to obtain. 
		Reference \cite{pcsbl} derived a  sub-optimal closed-form solution for a similar optimization by examining its optimality condition. We also derive a closed form sub-optimal solution by examining the optimality condition of \eqref{eq:max_surrogate}, and the properties of support vector $\mathbf{u}_{m}$ in the next theorem, which is derived in Appendix \ref{app:A}}. 
	\begin{theorem}\label{theorem:opt}
	The  optimal value of $\alpha_m^{NZ}$ and $\alpha_m^Z$ parameters of matrix $\mathbf{A}$ i.e.,  $(\alpha_m^{NZ})^*$ and $(\alpha_m^{Z})^*$ is 
	\begin{align}\label{eq:A_range}
		(\alpha_m^{NZ})^*&\in\bigg[\frac{c}{\sum_{n=1}^{NK}u_{nm}e_{nm}+d}, \frac{c+\sum_{n=1}^{NK}u_{nm}}{\sum_{n=1}^{NK}u_{nm}e_{nm}+d}\bigg) \forall m, \\ \nonumber
		(\alpha_m^{NZ})^*&\in\bigg[\frac{c}{\sum_{n=1}^{NK}(1-u_{nm})e_{nm}+d}, \frac{c+\sum_{n=1}^{NK}(1-u_{nm})}{\sum_{n=1}^{NK}(1-u_{nm})e_{nm}+d}\bigg) \forall m.
	\end{align}
	where and $ e_{nm} = |\mu_{nm}|^2+{\tau_{nm}}$. The scalar $\mu_{nm}$is the $n$th entry of the mean vector $\boldsymbol{\mu}_m$,  while  $\tau_{nm}$  is the $(n,n)$th entry of the covariance matrix $\boldsymbol{\Sigma}_m$. We note that  $(\alpha_{m}^i)^*$ lies in the range given in \eqref{eq:A_range}. Any value in this range is a suboptimal solution of \eqref{eq:max_surrogate}.  We chose the lower bound without any loss of generality 
	\begin{align}\label{eq:precisionA}
		\alpha_{m}^{NZ}=\frac{c}{\sum_{n=1}^{NK}u_{nm}e_{nm}+d} \text{ and } \alpha_{m}^{Z}=\frac{c}{\sum_{n=1}^{NK}(1-u_{nm})e_{nm}+d}\forall m.
	\end{align} 
	
\end{theorem}

$\bullet$ \textbf{Estimation of binary support matrix $\mathbf{U}$ of the cascaded VAD channel $\tilde{\mathbf{H}}$:}
For the prior in \eqref{eq:prior}, the prior precision of elements of $\tilde{\mathbf{h}}_m$, which is the $m$th column of  $\tilde{\mathbf{H}}$, depends on its support $\mathbf{u}_m$, whose $(n,m)$th entry $u_{nm}$ satisfies $u_{nm} = \mathbb{I}(\tilde{h}_{nm} \neq 0)$.
We now aim to detect this support with a high accuracy using the index-wise log-likelihood ratio test (LLRT) \cite{llrpaper}.
The authors in \cite{cp-sbl} derived the following LLRT for the $(n,m)$th element of $\tilde{\mathbf{H}}$:
\begin{align}
\label{eq:column_llrt}
\frac{((\tilde{\boldsymbol{\theta}}_{m})^H (\sigma^2\mathbf{I}_Q+\tilde{\boldsymbol{\Theta}} \text{diag}( \boldsymbol{\gamma}_{\backslash n,m} ) \tilde{\boldsymbol{\Theta}}^H)^{-1}\tilde{\mathbf{y}}_m)^2}{\tilde{\boldsymbol{\theta}}_{m}^H(\sigma^2\mathbf{I}_Q+\tilde{\boldsymbol{\Theta}} \text{diag}(  \boldsymbol{\gamma}_{\backslash n,m} )  \tilde{\boldsymbol{\Theta}}^H)^{-1}\tilde{\boldsymbol{\theta}}_{m}} \geqslant \bar{\theta}.
\end{align}
Here $\tilde{\boldsymbol{\theta}}_{m} \in \mathbb{C}^{Q \times 1}$ is the $m$th column of the pilot matrix $\tilde{\boldsymbol{\Theta}}$ and \sloppy{$ \boldsymbol{\gamma}_{\backslash n,m} = [\gamma_{1,m}, \cdots,\gamma_{n-1,m}, 0, \gamma_{n+1,m},\cdots, \gamma_{NK,m} ]^T$}. For a desired probability of false alarm $\epsilon \in [0,1]$, i.e., the probability that an element is zero but declared non-zero, the threshold $\bar{\theta}=\bigg(\mathcal{Q}^{-1}\big(\frac{\epsilon }{2}\big)\bigg)^2$. The term $\mathcal{Q}$ is the standard $\mathcal{Q}$-function \cite{llrpaper}.
The binary support matrix $\mathbf{U} \in \left\{0,1\right\}$ is estimated using the index-wise LLRTs in \eqref{eq:column_llrt} 
as:  $\hat{u}_{nm}=1$ if \eqref{eq:column_llrt} is true, and $0$ otherwise.  The estimate is denoted as $\hat{\mathbf{U}}$ with its $(n,m)$ entry being $\hat u_{nm}$. 
This completes the M-step of the proposed SM-SBL algorithm. The steps of SM-SBL are next summarized in Algorithm \ref{algo:CPSBL}.
\normalem
\begin{algorithm}
\DontPrintSemicolon 
\scriptsize
\KwIn{$\{\tilde{\mathbf{Y}}\}$, $\tilde{\boldsymbol{\Theta}}$ and initialize $t = 1, \gamma_{nm}^{(0)}=10^{-2},\forall n =1, \hdots, NK, \forall m=1, \hdots, M$, $\mathbf{U}^{(0)}=\mathbf{0}_{NK \times NK}$.}
\While{$t \leq T_{\text{max}}$ \& $\Delta \geq \eta $ }{
	
	\textbf{E-step}: Update diagonal elements of posterior covariance $\boldsymbol{\Sigma}_m$, and the mean $\boldsymbol{\mu}_m, m=1, \hdots, M,$  using \eqref{eq:W_cov}, respectively.\\
	
	\textbf{M-step}: Update precision parameter matrix $\alpha_{m}^{NZ} \text{ and } \alpha_{m}^{NZ} , \forall m=1, \hdots, M,$   according to \eqref{eq:precisionA}.\\
	\hspace{27pt} Update  binary support $ \hat{u}_{nm}, \forall n =1, \hdots, N, m=1, \hdots, M,$ according to  \eqref{eq:column_llrt}.\\
	$t \leftarrow t+1$, 			$\Delta= \sum_{m=1}^M({|| \boldsymbol{\mu}_m^{(t-1)}-\boldsymbol{\mu}_m^{(t)}||_2}/{||\boldsymbol{\mu}_m^{(t-1)}||_2})$.\\
}\Return{$\widehat{\tilde{\mathbf{H}}} \leftarrow [\boldsymbol{\mu}_1, \hdots, \boldsymbol{\mu}_M ]^{(t)}, \hat{\mathbf{U}}^{(t)}$}\;
\caption{Structured Mean field SBL (SM-SBL) algorithm.}\label{algo:CPSBL}
\end{algorithm}

\subsection{Factorized mean field sparse Bayesian learning algorithm}

The SM-SBL algorithm proposed above has a high time complextiy as the covariance matrix computation in \eqref{eq:W_cov} inverts an $NK\times NK$ \textit{full covariance matrix}. This has a time complexity of  $\mathcal{O}(N^3K^3)$. For a practical RIS-aided wireless system, with a large number of RIS elements $N$, the SM-SBL algorithm will  have a radically high time complexity. The author in \cite{fmf-sbl} proposed the fast mean field SBL (FMF-SBL) algorithm based on the fast mean field approximation \cite{fmf-sbl}. We now use the fast mean field approximation from \cite{fmf-sbl} to develop a faster version of SM-SBL, referred to as the factorized mean field SBL (FM-SBL) algorithm, to reduce its time complexity. The authors in \cite{fmf-sbl} have derived the updates for a simple uni-variate Gaussian prior which does not capture the complete sparsity structure of the joint cascaded VAD channel $\tilde{\mathbf{H}}$. They have used a real Gaussian distribution which is not applicable in RIS-assisted mmWave systems because the channel in these systems is complex. To address these issues we derive the updates using the proposed column-wise coupled prior which as discussed earlier captures the complete sparsity of $\tilde{\mathbf{H}}$. We also use complex Gaussian distribution to correctly model the channel.  We will show later that reduction in time complexity of the FM-SBL algorithm is achieved without sacrificing channel estimation accuracy.We need to discuss as to how our algoirhm is not a straight forward application of the algorithm in \cite{fmf-sbl}

The FM-SBL algorithm assumes a fully-factorized version of the approximating posterior distribution of \eqref{eq:approximating_posterior_form} i.e.,
\begin{align}\label{eq:factorized_posterior}
q(\tilde{\mathbf{h}}_m) = \prod_{n=1}^{NK} \mathcal{CN}(\tilde{h}_{nm}\vert \mu_{nm},\tau_{nm}) = \mathcal{CN}(\tilde{\mathbf{h}}_m\vert \boldsymbol{\mu_{m}},\text{diag}(\boldsymbol{\tau}_m)).
\end{align}
The variational parameters $\boldsymbol{\mu_{m}}\text{ and }\boldsymbol{\tau}_m$ respectively are the mean and marginal
variance of the factorized normal distribution. The E-step of the FM-SBL algorithm updates $\boldsymbol{\mu_{m}}\text{ and }\boldsymbol{\tau}_m$. The E-step also uses the Lipschitz inequality to bound the factorized ELBO. These two steps reduce the complexity of update steps of the variational parameters. The M-step of the FM-SBL is exactly same as the SM-SBL algorithm.


\subsubsection{\underline{E-step of the FM-SBL algorithm}}

The E-step again assumes that $\mathbf{A}$ and $\mathbf{U}$ are fixed at their current values and estimates the posterior distribution. The factorized ELBO expression is derived in the following lemma, whose proof is relegated to Appendix \ref{app:lambda_structured}. For notational simplicity we remove $\mathbf{A}$ and $\mathbf{U}$ from the ELBO expression on left hand side since they are fixed in the E-step.
\begin{lemma}
The ELBO for the fully-factorized approximating posterior distribution of \eqref{eq:factorized_posterior} is defined as follows:
\begin{align}\label{eq:ELBO_factorized}
	\mathcal{L}_F[\boldsymbol{\mu}_m,\boldsymbol{\tau}_m] &= -\frac{1}{\sigma^2}\lambda(\boldsymbol{\mu}_m,\boldsymbol{\tau}_m)-\sum_{n=1}^{NK}\beta_{n}(\boldsymbol{\mu}_m,\boldsymbol{\tau}_m)(\gamma_{nm})^{-1} + H(\boldsymbol{\tau}_m)  \\ \nonumber
	&=-\frac{1}{\sigma^2}||\tilde{\mathbf{y}}_m-\tilde{\boldsymbol{\Theta}}\boldsymbol{\mu}_m||^{2}_2-\frac{1}{\sigma^2}\mathbf{a}^{T}\boldsymbol{\tau}_m-\sum_{n=1}^{NK}(\vert\mu_{nm}\vert^2+\tau_{nm})(\gamma_{nm})^{-1} +  \sum_{n=1}^{NK} \ln \tau_{nm}. 
\end{align}
Here $  \lambda(\boldsymbol{\mu}_m,\boldsymbol{\tau}_m) = 
||\tilde{\mathbf{y}}_m-\tilde{\boldsymbol{\Theta}}\boldsymbol{\mu}_m||^{2}_2+\mathbf{a}^{T}\boldsymbol{\tau}_m, 
\beta_{n}(\boldsymbol{\mu}_m,\boldsymbol{\tau}_m)  = |\mu_{nm}|^{2} + \tau_{nm}$,  and 
$H(\boldsymbol{\tau}_m) =  \sum_{n=1}^{NK} \ln \tau_{nm}$. Also $\mathbf{a} = \text{diag}(\tilde{\boldsymbol{\Theta}}^{H}\tilde{\boldsymbol{\Theta}})$. 
\end{lemma}
As discussed in \cite{fmf-sbl}, the presence of the $l_2$ norm in \eqref{eq:ELBO_factorized} leads to  prohibitively expensive updates to $\boldsymbol{\mu}_m$. To obtain faster updates, we need to bound $||\mathbf{\tilde{y}}_m-\tilde{\boldsymbol{\Theta}}\boldsymbol{\mu}_m||_2^2$ such that it leads to less expensive updates. This objective is achieved by using the majorization framework. Instead of trying to maximize the exact function, we optimize a lower bound on the function. We apply the following lemma to obtain a majorizing function for $\lambda(\boldsymbol{\mu}_m,\boldsymbol{\tau}_m)$ of \eqref{eq:ELBO_factorized}.

\begin{lemma}
For any continuously differentiable function $h$ : 
$\mathcal{R}^{n} \rightarrow \mathcal{R}$ with an $L_{h}$ Lipschitz gradient, the inequality is applicable \cite{fmf-sbl}
\begin{align}\label{eq:lipschitz_inequality}
	h(\mathbf{x}) &\leq g(\mathbf{x} ; \mathbf{z}) 
	\triangleq h(\mathbf{z}) + (\mathbf{x}-\mathbf{z})^{H} \nabla_{x}h(\mathbf{z}) + \frac{L}{2} ||\mathbf{x}-\mathbf{z}||_2^{2}.
\end{align}
\end{lemma}
We apply this lemma by taking $h(\boldsymbol{\mu}_m) = ||\tilde{\mathbf{y}}_m - \tilde{\boldsymbol{\Theta}}\boldsymbol{\mu}_m||_2^{2}$. This leads to the following majorizing function for $\lambda(\boldsymbol{\mu}_m,\boldsymbol{\gamma}_m)$:
\begin{align} 
\label{eq:mm_lambda}
\Lambda(\boldsymbol{\mu}_m,\boldsymbol{\tau}_m;\boldsymbol{\delta}) & =  ||\tilde{\mathbf{y}}_m - \tilde{\boldsymbol{\Theta}}\boldsymbol{\delta}||_2^{2} + 2(\boldsymbol{\mu}_m - \boldsymbol{\delta})^{H}\tilde{\boldsymbol{\Theta}}^{H}(\tilde{\boldsymbol{\Theta}}\boldsymbol{\delta} - \tilde{\mathbf{y}}_m) \nonumber + \frac{L}{2}||\boldsymbol{\mu}_m - \boldsymbol{\delta}||^{2} + 
\mathbf{a}^{T}\boldsymbol{\tau}_m.
\end{align}

Substituting this into ELBO of \eqref{eq:common_ELBO}, we get the lower bound for  $\mathcal{L}_F[\boldsymbol{\mu}_m,\boldsymbol{\tau}_m]$, which is denoted as $\mathcal{L}_{FMF}[\boldsymbol{\mu}_m,\boldsymbol{\tau}_m;\boldsymbol{\delta}]$:
\begin{align}
\mathcal{L}_{FMF}[\boldsymbol{\mu}_m,\boldsymbol{\tau}_m;\boldsymbol{\delta}] &=
-\frac{1}{\sigma^2}\Lambda(\boldsymbol{\mu}_m,\boldsymbol{\tau}_m;\boldsymbol{\delta})-\sum_{n=1}^{NK}\beta_n(\boldsymbol{\mu}_m,\boldsymbol\tau_{m})(\gamma_{nm})^{-1} + H(\boldsymbol{\tau}_m). 
\end{align}
By construction, $\mathcal{L}_F[\boldsymbol{\mu}_m,\boldsymbol{\tau}_m] \geq \mathcal{L}_{FMF}[\boldsymbol{\mu}_m,\boldsymbol{\tau}_m;\boldsymbol{\delta}], \forall \boldsymbol{\delta}$ and $\mathcal{L}_F[\boldsymbol{\mu}_m,\boldsymbol{\tau}_m] = \mathcal{L}_{FMF}[\boldsymbol{\mu}_m,\boldsymbol{\tau}_m;\boldsymbol{\mu_m}]$.
The updates of $\boldsymbol{\mu}_m$ and $\tau_{nm}$ in the FM-SBL algorithm are derived by applying the first order stationarity conditions \cite{fmf-sbl}. To apply the stationarity conditions, we first calculate the gradient of $\mathcal{L}_{FMF}[\boldsymbol{\mu}_m,\boldsymbol{\tau}_m]$ with respect to $\boldsymbol{\mu}_m$ and $\tau_{nm}$, which are given as follows:
\begin{align}\label{eq:CPFMF_gradients}
\nabla_{\tau_{nm}} \mathcal{L}_{FMF}[\boldsymbol{\mu}_m,\boldsymbol{\tau}_m] &= \frac{ a_{n}}{\sigma^2} + (\gamma_{nm})^{-1} - \frac{1}{\tau_{nm}}, \\ \nonumber
\nabla_{\boldsymbol{\mu}_m} \mathcal{L}_{FMF}[\boldsymbol{\mu}_m,\boldsymbol{\tau}_m] &= \frac{2}{\sigma^2} \tilde{\boldsymbol{\Theta}}^{H}(\tilde{\boldsymbol{\Theta}}\boldsymbol{\delta}-\tilde{\mathbf{y}}_m) + \frac{ L}{2\sigma^2}*2(\boldsymbol{\mu}_m-\boldsymbol{\delta}) + 2 \; \boldsymbol{\Xi} \boldsymbol{\mu}_m.
\end{align}

By applying the first order stationarity condition, we get the following updates for the mean and covariance matrix using the local ascent approach \cite{fmf-sbl}:
\begin{align}
&\boldsymbol{\mu}_m^{(t+1)} = \mathbf{D}^{(t)} \boldsymbol{\zeta}^{(t)},   \text { where, }
\boldsymbol{\zeta}^{(t)} = \frac{1}{\sigma^2}\left(\frac{L}{2}\boldsymbol{\mu}_m^{(t)}-\tilde{\boldsymbol{\Theta}}^{H}(\tilde{\boldsymbol{\Theta}}\boldsymbol{\mu}_m^{(t)}-\tilde{\mathbf{y}}_m) \right) \text{ and }
\mathbf{D}^{(t)} = \left( \frac{L}{2\sigma^2}\mathbf{I} + \boldsymbol{\Xi} \right)^{-1}, \notag \\
\label{eq:varianceFMF}
&\text{ and } 
\boldsymbol{\tau}_m^{(t+1)} = (\frac{1}{\sigma^2} \mathbf{a} + \boldsymbol{\alpha}_m)^{-1}.
\end{align}
In order to obtain \eqref{eq:varianceFMF}, the substitution $\boldsymbol{\delta} = \boldsymbol{\mu}_m^{(t)}$ was made.
Here $\boldsymbol{\Xi}$ is the precision matrix, $\boldsymbol{\alpha}_m$ is the prior precision vector with $\boldsymbol{\Xi} = \text{diag}(\boldsymbol{\alpha}_m)$ and $\mathbf{a} = \text{diag}(\boldsymbol{\tilde{\Theta}}^{H}\boldsymbol{\tilde{\Theta}})$.
Here $\boldsymbol{\alpha}_{m}$ is the $m$th column of the prescision matrix $\mathbf{A}$ and $L = 2\lambda_{max}(\tilde{\boldsymbol{\Theta}}^{H}\tilde{\boldsymbol{\Theta}})$.
While deriving the mean and variance updates, we have assumed that the precision matrix $\mathbf{A}$ and the binary support matrix $\mathbf{U}$ are known. These hyperparameters are estimated similar to the SM-SBL algorithm. The precision updates are calculated by maximizing the expected value of the complete data log likelihood and the support is estimated from the LLRT test.

\subsubsection{{M-step of the FM-SBL algorithm}}
The M-step is the same as the SM-SBL algorithm. The precision matrix $\mathbf{A}$ is updated according to \eqref{eq:precisionA}. The binary support matrix is updated according to \eqref{eq:column_llrt}. The proposed FM-SBL algorithm, thus, solves the time complexity issues of the SM-SBL algorithm by using a factorized approximating posterior and also using the Lipschtiz inequality to bound the factorized ELBO. The steps of FM-SBL are next summarized in Algorithm \ref{algo:CPFMFSBL}.
\begin{algorithm}[htbp]
\scriptsize
\DontPrintSemicolon 
\KwIn{$\{\tilde{\mathbf{Y}}\}$, $\tilde{\boldsymbol{\Theta}}$.}
\SetKwInOut{Initialization}{Intialize}
\Initialization{$t = 1, \gamma_{nm}^{(0)}=10^{-2},\forall n =1, \hdots, N, \forall m=1, \hdots, M$, $\mathbf{B}_{m}^{(0)}=\frac{\mathbf{1}_{NK \times NK}}{NK}, \forall m =1, \hdots, M$.}
\While{$(t \leq T_{max} $ \& $\Delta_m > \eta_1 )$}{
	\SetKwInOut{Initialization}{Intialize} $m = 1$.\\
	\While{$(m \leq M$)}{
		\textbf{E-step}: Update diagonal elements of posterior covariance $\boldsymbol{\Sigma}_m$, and the mean $\boldsymbol{\mu}_m,$  using \eqref{eq:varianceFMF} and \eqref{eq:varianceFMF}, respectively.\\
		\textbf{M-step}: Update parameter matrix $\alpha_{m}^{NZ} \text{ and } \alpha_{m}^{Z},$   according to \eqref{eq:precisionA}.\\
		\hspace{29pt}Update  binary support $ \hat{u}_{nm}, \forall n =1, \hdots, N$ according to  \eqref{eq:column_llrt}.\\
		$m \leftarrow m+1$ \\
	}
	$t \leftarrow t+1$, 			$\Delta= \sum_{m=1}^M({|| \boldsymbol{\mu}_m^{(t-1)}-\boldsymbol{\mu}_m^{(t)}||_2}/{||\boldsymbol{\mu}_m^{(t-1)}||_2})$.\\
}
\Return{$\widehat{\tilde{\mathbf{H}}} \leftarrow [\boldsymbol{\mu}_1, \hdots, \boldsymbol{\mu}_M ]^{(t)}, \hat{\mathbf{U}}^{(t)}$.}\;
\caption{Factorized mean field SBL (FM-SBL) algorithm.}\label{algo:CPFMFSBL}
\end{algorithm}

\subsection{Complexity analysis of SM-SBL and FM-SBL}
We now show the per-iteration time complexity of the proposed SM-SBL and FM-SBL algorithms in Table \ref{t1} and Table \ref{t2}, respectively. The computation complexity of calculating the covariance matrix $\boldsymbol{\Sigma}_m$ in \eqref{eq:W_cov}, by using the Woodbury identity \cite{sbl} $\boldsymbol{\Sigma}_m =\boldsymbol{\Upsilon}_{m} - \boldsymbol{\Upsilon}_{m} \boldsymbol{\tilde{\Theta}}^H (\sigma^{2} \mathbf{I}_Q + \boldsymbol{\tilde{\Theta}} \boldsymbol{\Upsilon}_{m} \boldsymbol{\tilde{\Theta}}^H )^{-1} \boldsymbol{\tilde{\Theta}} \boldsymbol{\Upsilon}_{m}$, is reduced  from $\mathcal{O}(N^3K^3)$ to $\mathcal{O}(Q^3)$. The FM-SBL algorithm, as derived in \eqref{eq:varianceFMF}, updates the variance of each element independently. The mean update for FM-SBL in \eqref{eq:varianceFMF}  inverts a diagonal matrix with $\mathcal{O}(NK)$ complexity. The mean and variance updates of FM-SBL algorithms do not invert a full matrix, and hence its is much faster than the SM-SBL algorithm. The index-wise LLRTs step to update $\hat{\mathbf{u}}_m$ for both algorithms has the highest complexity.

\begin{table*}[htbp]
\centering
{
	\begin{subtable}{0.5\linewidth}
		\centering 
		\begin{tabular}{|c|c|}
			\hline
			Variable updates & proposed SM-SBL
			\\ \hline
			${\boldsymbol{\mu}}_{m},{\boldsymbol{\Sigma}}_{m}, \forall m$  &$\mathcal{O}((Q^3+(NK)^2)M)$     \\
			$\boldsymbol{\alpha}_{m},\forall m$& $\mathcal{O}(NKM)$   \\
			$\hat{u}_{nm}, \forall n,m$&$\mathcal{O}(NKQ^3M+ NKM)$    \\ \hline
		\end{tabular}\caption{}\label{t1}
	\end{subtable}
	\begin{subtable}{0.5\linewidth}
		\centering 
		\begin{tabular}{|c|c|}
			\hline
			Variable updates & proposed FM-SBL 
			\\ \hline
			${\boldsymbol{\mu}}_{m},{\boldsymbol{\Sigma}}_{m}$  &$\mathcal{O}((NK+(NK)^2)M)$     \\
			$\boldsymbol{\alpha}_{m},\forall m$& $\mathcal{O}(NKM)$   \\
			$\hat{u}_{nm}, \forall n,m$&$\mathcal{O}(NKQ^3M+ NKM)$    \\ \hline
		\end{tabular}\caption{}\label{t2}
\end{subtable}}
\caption{(a) Per iteration complexity of SM-SBL. (b) Per iteration complexity of FM-SBL.}
\end{table*}

\section{Numerical results}
We investigate the performance of the proposed SM-SBL and FM-SBL algorithm in this section for channel estimation in RIS-assisted mmWave systems. We consider the following metrics to compare the performance, (i) NMSE, for channel estimation which is defined as $||\mathbf{\hat{\tilde{H}}}-\mathbf{\tilde{H}}||_F^2/||\mathbf{\tilde{H}}||_F^2$, (ii) normalized support error rate (NSER), which measures the error in the estimated support and is defined as $||\mathbf{\hat{U}}-\mathbf{U}||_F^2/||\mathbf{U}||_F^2$, (iii) KL divergence which measures the difference between the true posterior distribution and the estimated approximate posterior distribution. It is defined as, $KL(p||q) = -\mathbb{E}_{q}\left[\log \frac{p}{q}\right]$, where $p$ is the true posterior distribution and $q$ is the approximate posterior distribution, (iv) sum-spectral efficiency (sum-SE) which describes the amount of data transmitted over a given spectrum. The BS precodes the UE data $\mathbf{s}\in\mathbb{C}^{K \times 1}$ by maximum ratio combining (MRC),
which for the kth UE is given as $\mathbf{v}^k = \frac{\hat{\mathbf{H}}^k\boldsymbol{\theta}}{\sqrt{\mathbb{E}[||\hat{\mathbf{H}}^k\boldsymbol{\theta}||^2]}}\in\mathbb{C}^{M \times 1}$. The signal for the $k$th UE as detected by the BS is
\begin{align} \nonumber
y^k = (\mathbf{v}^k)^H\mathbf{H}^k\boldsymbol{\theta}s^k + (\mathbf{v}^k)^H\sum_{n\neq k}\mathbf{H}^n\boldsymbol{\theta}s^n + (\mathbf{v}^k)^H\mathbf{w}^k
\end{align}
. The sum-SE (in bps/Hz) for the K UEs is defined as follows:
\begin{align} \nonumber
\text{SE} = \left(1-\frac{T_p}{T_c}\right)\sum_{k=1}^K\mathbb{E}\left[\log_2\left(1 +\frac{|(\mathbf{v}^k)^H\mathbf{H}^k\boldsymbol{\theta}|^2}{(\mathbf{v}^k)^H\sum_{n\neq k}\mathbf{H}^n\boldsymbol{\theta}+\sigma^2} \right)\right].
\end{align}
Here $T_p$ and $T_c$ denote the length of the pilots and the coherence interval, respectively,
and (v) finally we plot the run-time  which measures the difference in execution time between the proposed algorithms. These studies are performed by considering a BS with $M=64$ antennas and a UPA RIS with $N=128$ elements.  The total number of UEs is $K=4$. The number of paths between the BS and RIS is $L_C=20$, and the number of paths between the RIS and each UE is $L_R^k=40$. The elements of RIS reflecting matrix $\boldsymbol{\Theta}$ are selected uniformlly from $\{-\frac{1}{\sqrt{N}},\frac{1}{\sqrt{N}}\}$, similar to \cite{ris}. We also compare the performance of the proposed algorithms with the following algorithms: (i) DS-OMP \cite{ris}: doubly-structured sparsity based algorithm for channel estimation proposed in \cite{ris}. It uses the OMP algorithm by incorporating the sparsity structure of the cascaded VAD channel; (ii) SBL \cite{sbl}: solves the channel estimation using the SBL algorithm. It incorporates just the individual sparsities of the elements of cascaded VAD channel and not its doubly-structured sparsity. We also plot the performance of these algorithms against a lower bound, \textit{Oracle MMSE}, which knows the true support of the cascaded VAD channel and calculates the minimum mean squared error (MMSE) solution for this over-determined problem.

\begin{figure}[htbp]
\begin{subfigure}{0.32\linewidth}
	\includegraphics[width=\linewidth]{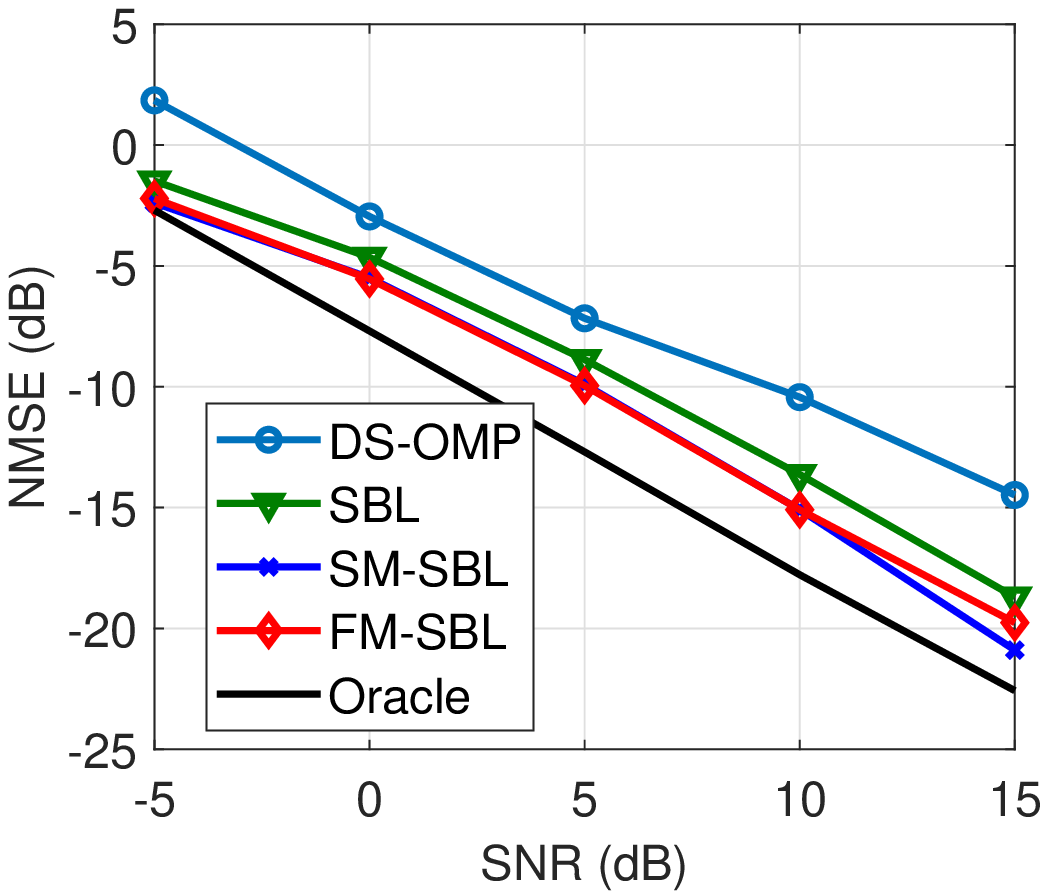}\vspace{-5pt}
	\caption{}\label{fig:SNR}
\end{subfigure}
\begin{subfigure}{0.32\linewidth}
\includegraphics[width=\linewidth]{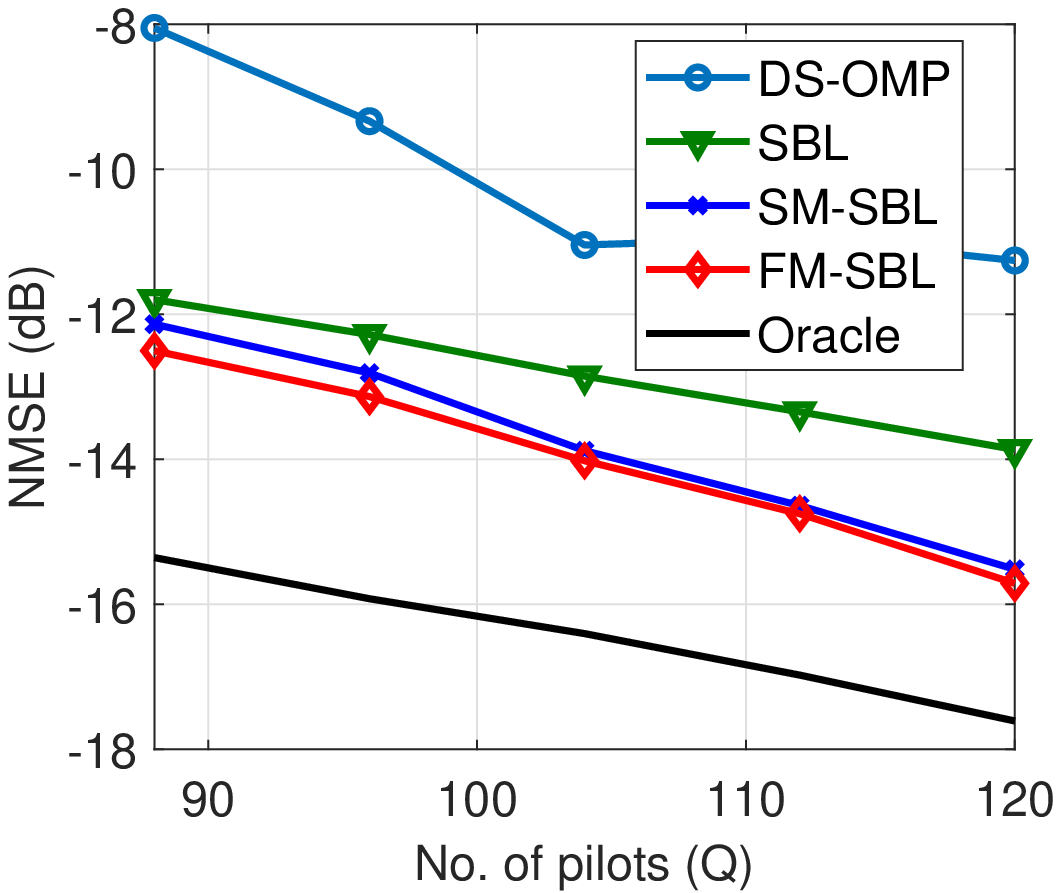}\vspace{-5pt}
\caption{}
\label{fig:QvsNMSE}
\end{subfigure}
\begin{subfigure}{0.32\linewidth}
\centering
\includegraphics[width=\linewidth]{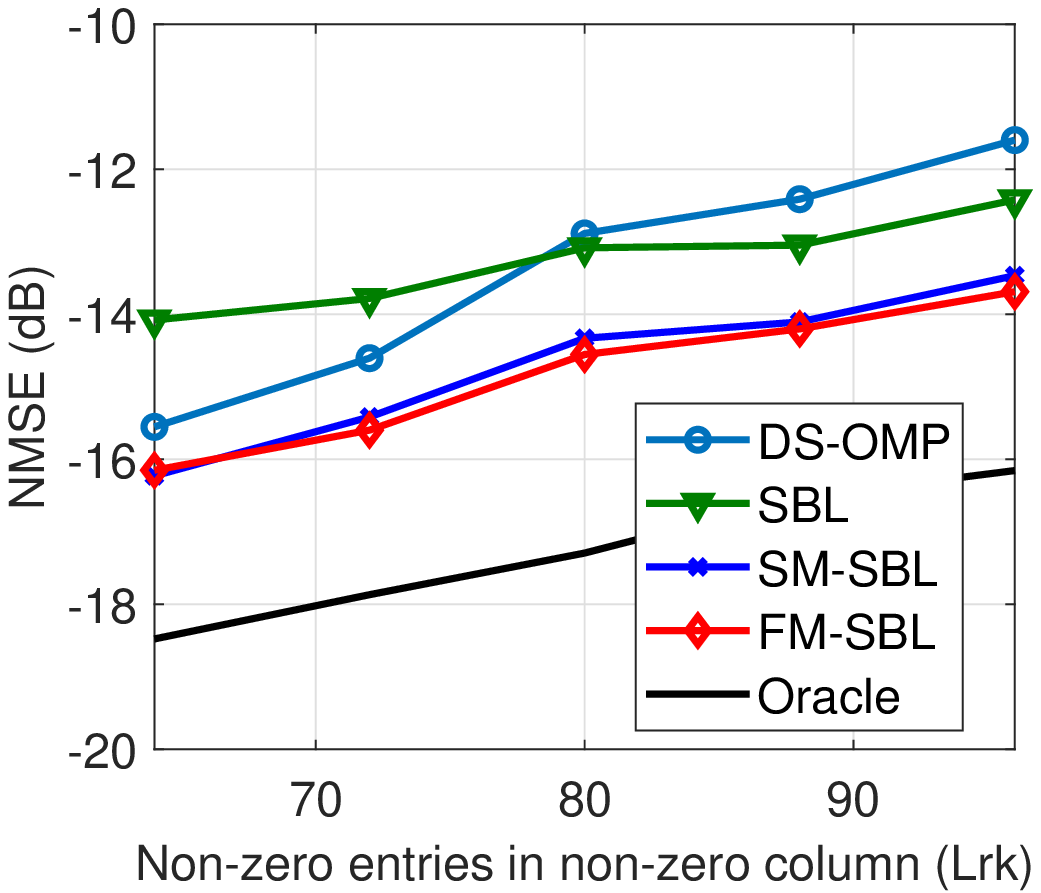}\vspace{-5pt}
\caption{}\label{fig:RowSparsity}
\end{subfigure}
\caption{(a)  NMSE  with SNR;  (b) NMSE with varying pilots, and; (c) NMSE with varying number sparsity level. }\vspace{-15pt}
\end{figure}
\textbf{NMSE Comparison:} We show in Fig. \ref{fig:SNR} the NMSE of various algorithms by varying the SNR $=P/\sigma^2=1/\sigma^2$, and  fixing number of pilots as $Q=100$. We see that the  proposed  SM-SBL and FM-SBL  algorithms have lower NMSE than the other algorithms. This is because they exploit the doubly-structured sparsity along with the individual sparsities of the cascaded  VAD channel $\tilde{\mathbf{H}}$. Their performance is also close to the Oracle MMSE.  The SM-SBL and FM-SBL algorithms give the same performance except at 15dB, where the latter performs slightly worse. This is because the factorized posterior assumption used in this algorithm, gives an underestimated variance \cite{save-amp}. This incorrect posterior variance issue is observed in other works as well \cite{save-amp}. The proposed algorithms also outperform the (i) SBL algorithm which does not exploit the doubly-structured sparsity across the UEs and (ii) DS-OMP algorithm which does not exploit the indvidual sparsities of each element of $\tilde{\mathbf{H}}$. It also requires the complete sparsity degree information of $\tilde{\mathbf{H}}$ which is obtained with high complexity \cite{gamp}.

We plot in Fig. \ref{fig:QvsNMSE} the NMSE by varying the number of pilots $Q$. We see that the NMSE  of all algorithms decreases with increase in $Q$. This is due to the increase in the number of observations. We also observe that the two proposed algorithms have the lowest NMSE for all values of $Q$. This is because the proposed algorithms exploit the entire sparsity structure of $\tilde{\mathbf{H}}$. DS-OMP which does not exploit the individual sparsity and SBL which does not exploit the doubly-structured sparsity have higher NMSE. The proposed algorithms require fewer number of pilots for channel estimation, thus help in reducing the pilot overhead. For example, both FM-SBL and SM-SBL yield a NMSE of $-14$dB with $Q=104$, while SBL requires $120$ pilots. The DS-OMP will require an even higher number of pilots.

\begin{figure}[htbp]
\centering
\begin{subfigure}{0.32\linewidth}
\includegraphics[width=\linewidth]{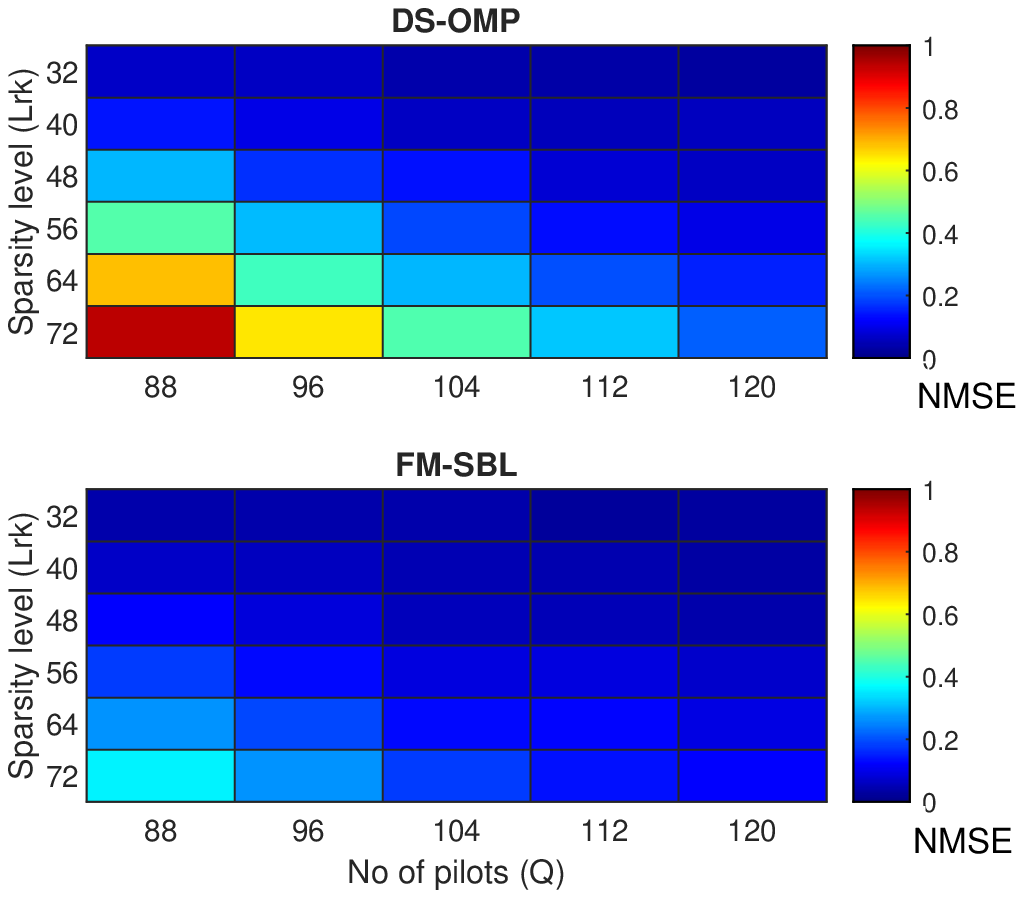}\vspace{-5pt}
\caption{}
\label{fig:PhasePlot}
\end{subfigure}
\begin{subfigure}{0.32\linewidth}
\includegraphics[width=\linewidth]{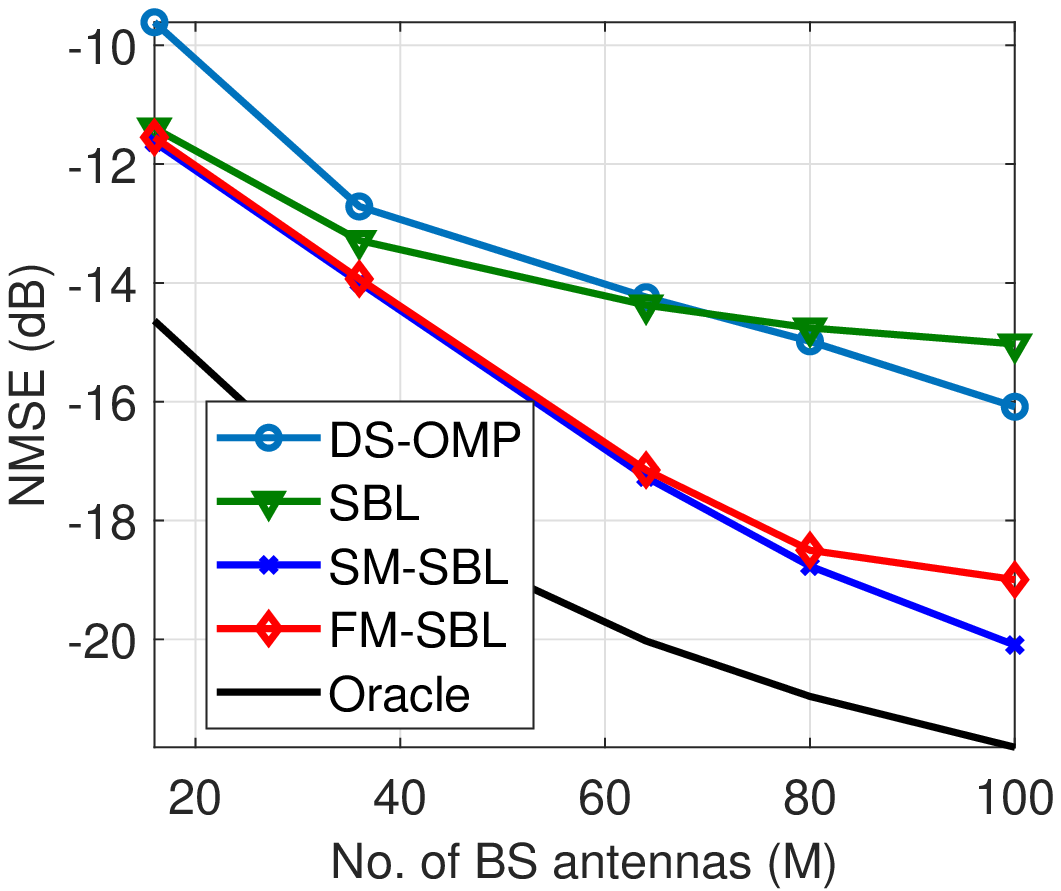}\vspace{-5pt}
\caption{}
\label{fig:MvsNMSE}
\end{subfigure}	
\begin{subfigure}{0.32\linewidth}
\includegraphics[width=\linewidth]{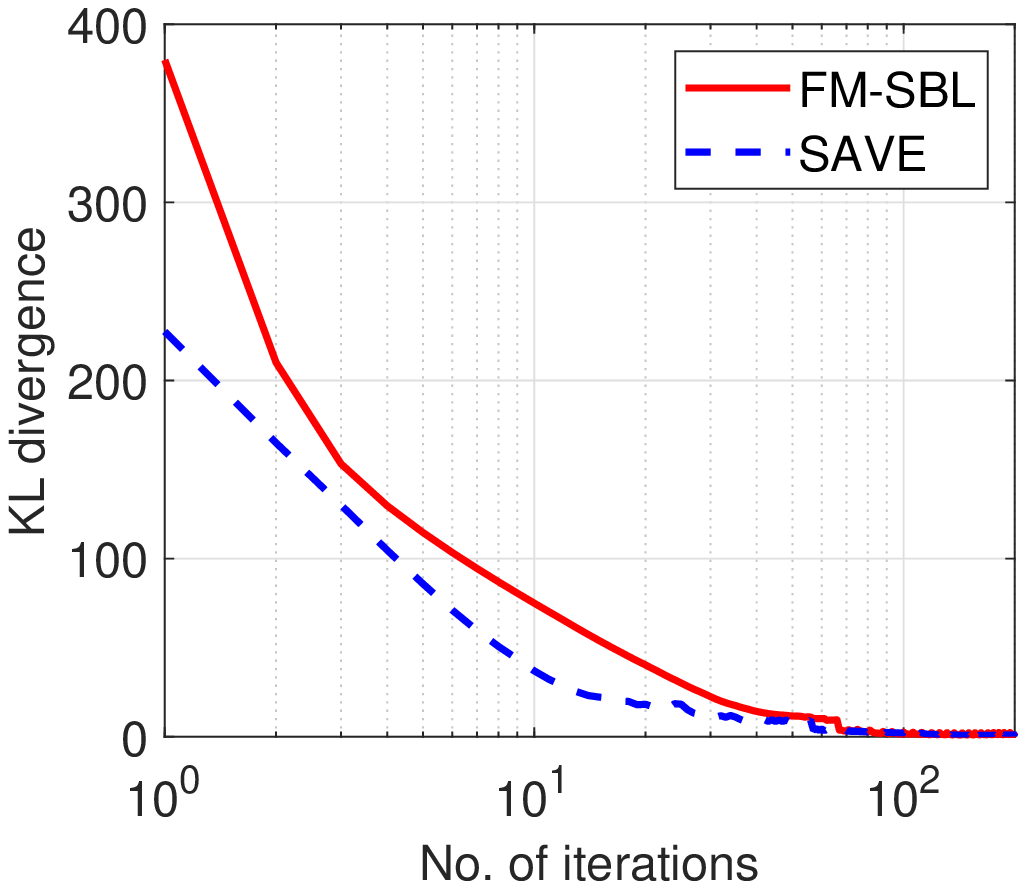}\vspace{-5pt}
\caption{}\label{fig:KL_div}
\end{subfigure}
\caption{(a) NMSE with varying pilots and sparsity level; (b) NMSE with varying measurements; and; (c) KL divergence.}\vspace{-0.4cm}
\end{figure}

We next plot in Fig. \ref{fig:RowSparsity} the NMSE values by varying the number of non-zero values inside each non-zero column of $\tilde{\mathbf{H}}$, denoted as $L_R^k$. This corresponds to varying the number of scatterers between the UE and the RIS. The SNR is fixed at $10$ dB. We first observe that the two proposed algorithms have the lowest NMSE. The DS-OMP algorithm performs just slightly worse than the proposed algorithms when the number of scatterers is low. This is because with low number of scatterers, the individual sparsities of each element is insignificant. Exploiting the doubly-structured sparsity gives the low NMSE values. With increase in the number of scatterers, keeping the number of non-zero columns $L_C$ and number of common entries in each non-zero column $L_R$ of $\tilde{\mathbf{H}}$ fixed, the individual sparsities become more significant. The NMSE of DS-OMP algorithm which does not epxploit the individual sparsity, hence, becomes worse with increase in $L_R^k$.  
We plot in Fig. \ref{fig:PhasePlot},  the NMSE values for the DS-OMP and FM-SBL algorithms by varying the number of pilots $Q$ and the number of scatterers between the RIS and UE which corresponds to varying $L_R^k$, as discussed earlier. We observe that for low $L_R^k$ values, both algorithms have low NMSE values. This is because there are sufficient number of pilots, and exploiting just the doubly-structured sparsity is enough. But for high $L_R^k$ values, the proposed FM-SBL algorithm has a much lower NMSE than the DS-OMP. This is because, as discussed earlier, it exploits both the individual and doubly-structured sparsities, which the DS-OMP does not. This makes the FM-SBL algorithm more robust to the change in number of scatters between the UE and the RIS. The proposed algorithms are thus a better choice for RIS channel estimation in different scattering environments.

We plot in Fig. \ref{fig:MvsNMSE} the NMSE values by varying the number of BS antennas $M$. We observe that for all algorithms the NMSE value decreases with the increase in $M$.  For a fixed number of scattering paths the sparsity in the channel remains constant. Increasing $M$, increases the number of observations, $QNKM$, without increasing sparsity. This reduces the NMSE values for all algorithms.  We once again observe that the proposed algorithms have the lowest NMSE, which is close to the Oracle lower bound. This shows the effectiveness of the proposed algorithms even for a large value of $M$, which is typical in the current 5G mmWave systems. For high $M$ values, the DS-OMP algorithm has a lower NMSE than the SBL algorithm. This is because with increase in $M$, the doubly-structured sparsity of $\tilde{\mathbf{H}}$ becomes significant, which the DS-OMP exploits better than SBL. 

\textbf{KL-divergence Comparison:} We plot in Fig. \ref{fig:KL_div} the KL-divergence against the number of iterations. As discussed earlier, the KL-divergence is a measure of distance between distributions. We compare the KL divergence between the SM-SBL posterior distribution and the posterior distributions estimated by the FM-SBL and SAVE algorithms. The SAVE algorithm uses the fully factorized posterior distribution of \eqref{eq:factorized_posterior}, and does not use the Lipschitz inequality of \eqref{eq:lipschitz_inequality} to bound the factorized ELBO. This plot helps us in understanding how the Lipschitz inequality affects the convergence properties of the proposed FM-SBL algorithm. We observe that the SAVE algorithm converges slightly faster than the FM-SBL algorithm. The KL-divergence for both algorithms, however, becomes close to zero with the increase in number of iterations. We conclude that the FM-SBL algorithm also converges to the SM-SBL posterior distribution, and Lipschitz inequality does not affect the final distribution to which the proposed FM-SBL algorithm converges. It marginally affects the convergence rate.
\begin{figure}[htbp]
\centering
\begin{subfigure}{0.32\linewidth}
\includegraphics[width=\linewidth]{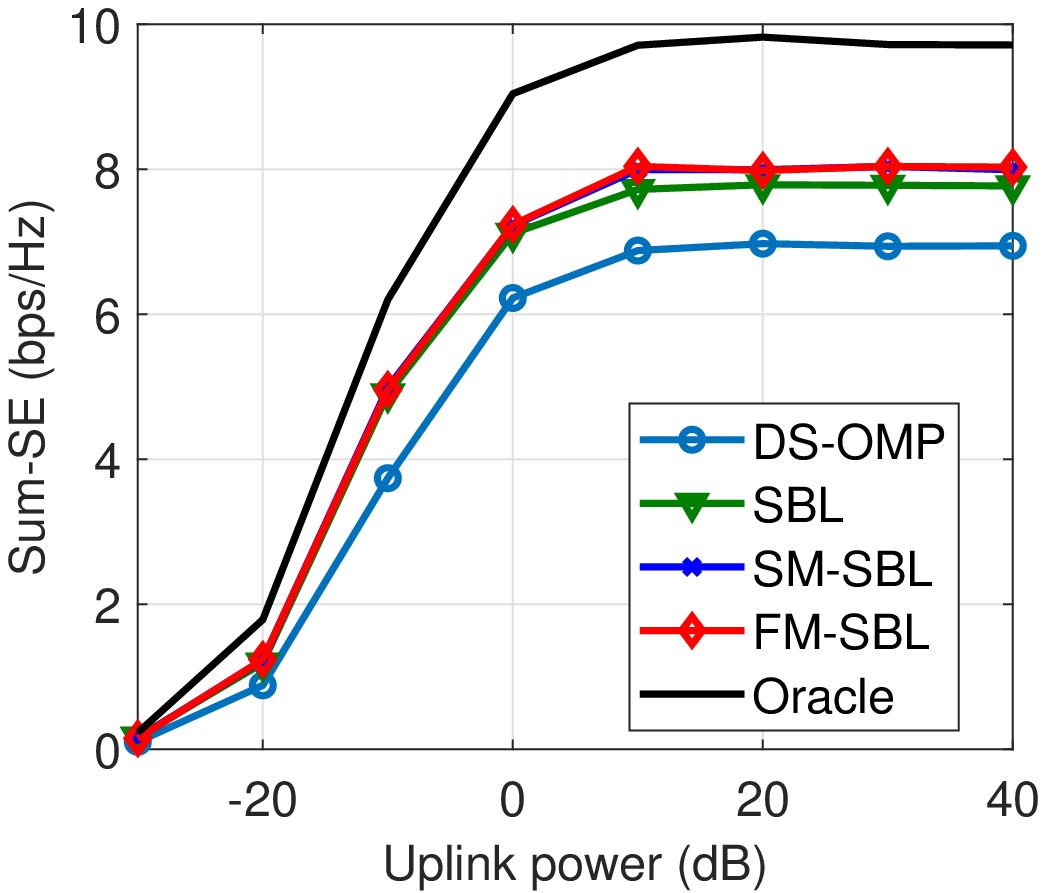}\vspace{-5pt}
\caption{}
\label{fig:Sumrate}
\end{subfigure}
\begin{subfigure}{0.32\linewidth}
\includegraphics[width=\linewidth]{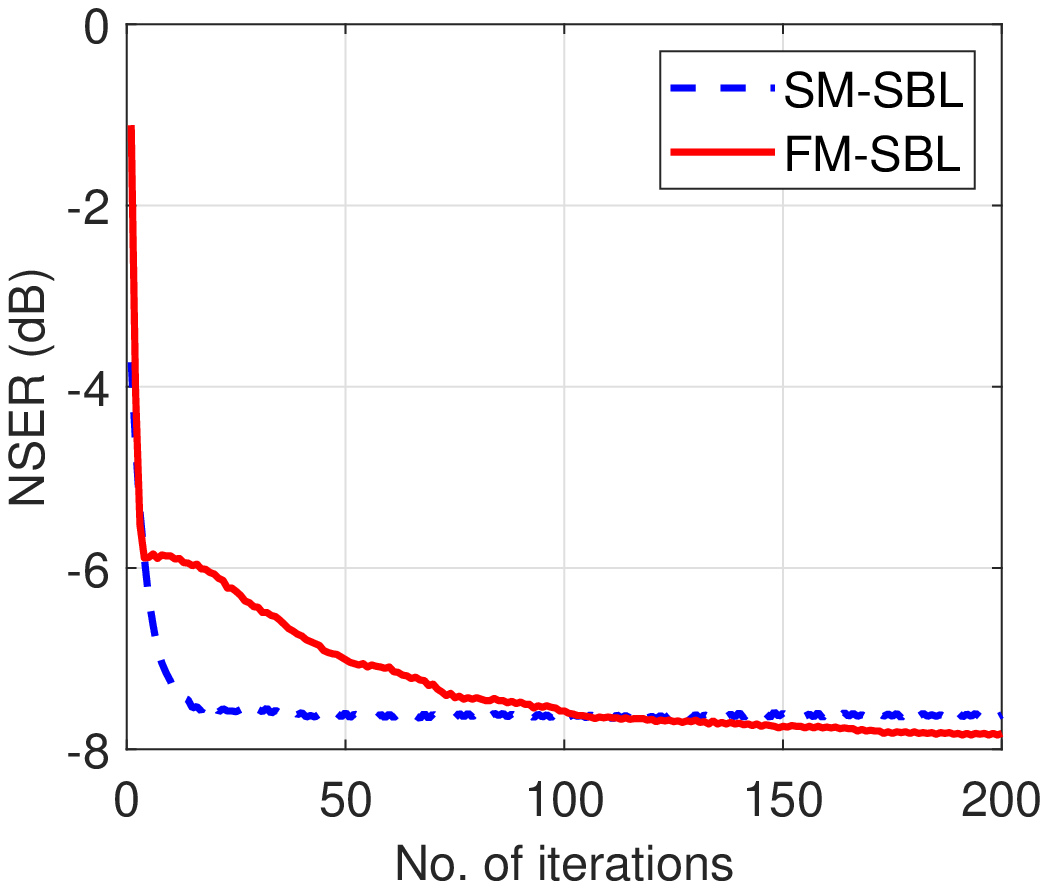}\vspace{-5pt}
\caption{}
\label{fig:NSER}
\end{subfigure}	
\begin{subfigure}{0.32\linewidth}
\includegraphics[width=\linewidth]{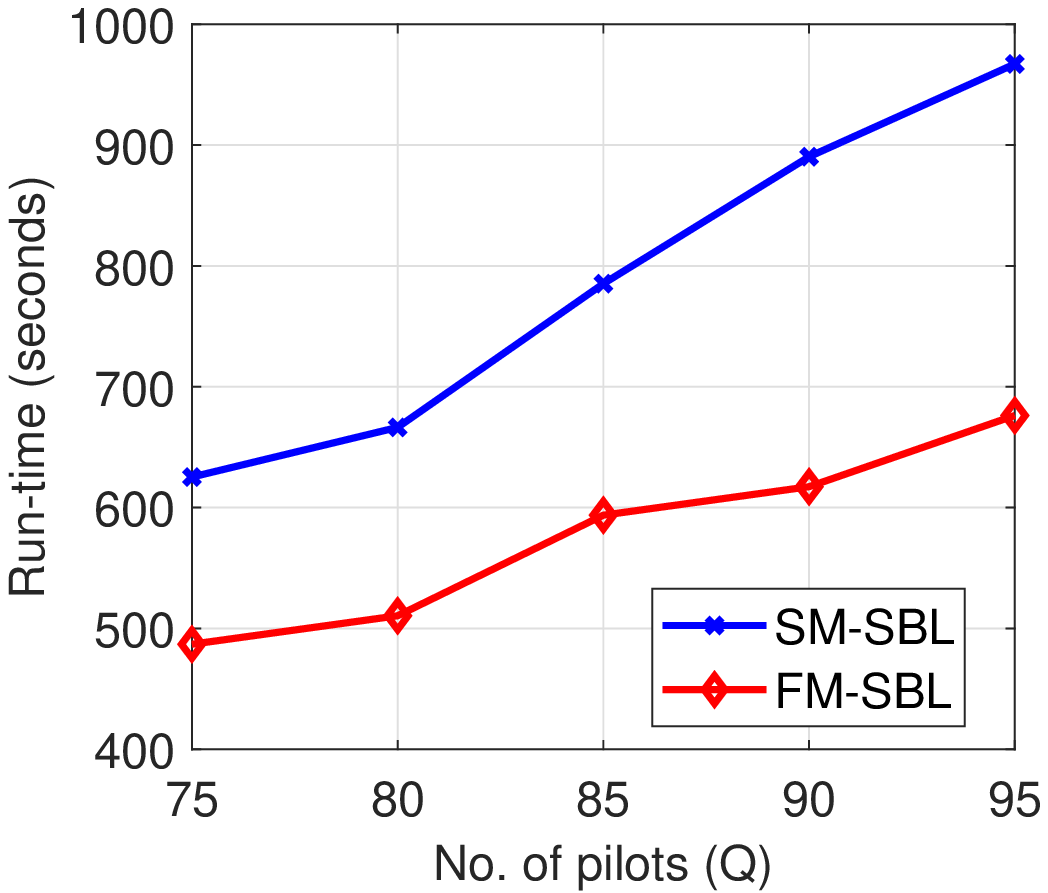}\vspace{-5pt}
\caption{}\label{fig:run-time}
\end{subfigure}
\caption{(a) Sum-SE comparision against Uplink Power; (b) NSER with Iterations; and; (c) Runtime.}\vspace{-0.4cm}
\end{figure}

\textbf{SE Comparison:} We in Fig. \ref{fig:Sumrate}, we compare the sum-SE of the proposed algorithms to the other existing ones. We plot the SE by varying the uplink power and by fixing the number of pilots at $Q=80$. The number of scatters between the BS and the RIS is set as $L_C=35$. The Oracle MMSE has the highest SE because its NMSE, as observed in Fig. \ref{fig:SNR}, is the lowest. The two proposed algorithms outperform the other algorithms in terms of sum-SE as well. This is due to their lower channel estimation NMSE values. For high values of uplink power, the sum-SE of all algorithms saturates. This is because of high multi-UEs interference. We also see that the SBL algorithm has higher sum-SE than the DS-OMP algorithm, this is due to a high number of scatterers between the BS and the RIS. This can also be verified from the sprsity plot of Fig. \ref{fig:RowSparsity}, where for a high number of scatterers, the SBL algorithm has a lower NMSE than the DS-OMP algorithm.

\textbf{NSER Comparison:} We compare in Fig. \ref{fig:NSER} the estimated support of $\tilde{\mathbf{H}}$ of the proposed algorithms against the true support at SNR=$15$dB. We observe that although the SM-SBL algorithm converges faster than the FM-SBL algorithm, the value at which it converged is same as the FM-SBL algorithm. The NSER values are low for the proposed algorithms. The support estimated by them is highly accurate. This again validates that the fact that the Lipschitz inequality does not affect the value at which the FM-SBL algorithm converges, it just affects the convergence rate.

\textbf{Runtime Comparison:} We finally in Fig. \ref{fig:run-time} plot the run-time of the proposed algorithms by varying the number of pilots $Q$. We observe that the FM-SBL algorithm has a significantly lower run-time than the SM-SBL algorithm. The DS-OMP algorithm will require much less time. This is expected as the proposed algorithms are iterative and do not require the knowledge of the number of UE-RIS and RIS-BS channel paths, which have to be provided as input for the DS-OMP. We also observe that run-time gap between the two proposed algorithms increase with the number of pilots $Q$. This makes the FM-SBL algorithm more suitable for mmWave systems, as its run-time scales at a lower rate with $Q$ than the SM-SBL algorithm.

\section{Conclusions}
We proposed two algorithms for CE in RIS assisted mmWave MIMO systems i.e., SM-SBL and FM-SBL, algorithms. Both these algorithms use the column wise coupled prior which is able to exploit the doubly-structured sparsity and the individual sparsity of the elements of angular cascaded channel. We showed through our numerical investigations that i) NMSE values of the proposed SM-SBL and FM-SBL algorithm are lower than the other state of the art algorithms and ii) The FM-SBL algorithm has lower runtime as compared to the SM-SBL algorithms.
\appendices
\section{}\label{app:lambda_structured}
\textbf{Derivation of Terms of Structured ELBO:} Here we derive the terms in the ELBO for the approximating posterior of \eqref{eq:approximating_posterior_form} which are denoted as $\lambda(\boldsymbol{\mu}_m,\boldsymbol{\Sigma}_m)$, $\beta_n(\boldsymbol{\mu}_m,\boldsymbol{\Sigma}_m)$ and $H(\boldsymbol{\Sigma}_m)$.
\begin{align}
\lambda(\boldsymbol{\mu}_m,\boldsymbol\Sigma_{m}) &= \langle||\tilde{\mathbf{y}}_m - \tilde{\boldsymbol{\Theta}}\tilde{\mathbf{h}}_m||_2^{2}\rangle = \langle (\tilde{\mathbf{y}}_m - \tilde{\boldsymbol{\Theta}}\tilde{\mathbf{h}}_m)^H(\tilde{\mathbf{y}}_m - \tilde{\boldsymbol{\Theta}}\tilde{\mathbf{h}}_m) \rangle \\ \nonumber
&\stackrel{(a)}{\propto} \langle \tilde{\mathbf{h}}_m^H\tilde{\boldsymbol{\Theta}}^H \tilde{\boldsymbol{\Theta}}\tilde{\mathbf{h}}_m\rangle - 2\langle\tilde{\mathbf{h}}_m\rangle^H\tilde{\boldsymbol{\Theta}}^H\tilde{\mathbf{y}}_m 
\stackrel{(b)}{=} \text{trace}(\langle \tilde{\mathbf{h}}_m^H\tilde{\boldsymbol{\Theta}}^H \tilde{\boldsymbol{\Theta}}\tilde{\mathbf{h}}_m\rangle) - 2\boldsymbol{\mu}_m^H\tilde{\boldsymbol{\Theta}}^H\tilde{\mathbf{y}}_m \\ \nonumber
&= \langle \text{trace}(\tilde{\mathbf{h}}_m^H\tilde{\boldsymbol{\Theta}}^H \tilde{\boldsymbol{\Theta}}\tilde{\mathbf{h}}_m)\rangle - 2\boldsymbol{\mu}_m^H\tilde{\boldsymbol{\Theta}}^H\tilde{\mathbf{y}}_m  
\stackrel{(c)}{=} \text{trace}(\tilde{\boldsymbol{\Theta}}^H \tilde{\boldsymbol{\Theta}}(\boldsymbol{\Sigma}_m+\boldsymbol{\mu}_m\boldsymbol{\mu}_m^H))- 2\boldsymbol{\mu}_m^H\tilde{\boldsymbol{\Theta}}^H\tilde{\mathbf{y}}_m \\ \nonumber 
&\stackrel{(d)}{=} ||\tilde{\mathbf{y}}_m-\tilde{\boldsymbol{\Theta}}\boldsymbol{\mu}_m||_2^2 + \text{trace}(\tilde{\boldsymbol{\Theta}}^H \tilde{\boldsymbol{\Theta}}\boldsymbol{\Sigma}_m)
\end{align}
Proportionality in (a) is obtained by just looking at the terms which are dependent on the variational parameters. Equality in (b) is obtained by replacing the expectation with the mean of the approximating distribution and the fact that the trace of a scalar is the same as the scalar value. Equality in (c) is obtained by using the cyclic property of the trace operation. The cyclic propert of trace states that $\text{trace}(\mathbf{AB})=\text{trace}(\mathbf{BA})$. Equality in (d) is obtained by the linearity of trace operation and rearranging the terms in form of the $l_2$ norm.
Now we derive the entropy term $H(\boldsymbol{\Sigma}_m)$,
\begin{align}
H(\boldsymbol{\Sigma}_m) &= - \mathbb{E}_q[\ln q(\tilde{\mathbf{h}}_{m})]
\stackrel{(a)}{=} \ln (\pi^n \det(\boldsymbol{\Sigma}_m)) \int q(\tilde{\mathbf{h}}_{m})(\tilde{\mathbf{h}}_{m}-\boldsymbol{\mu}_m)^H\boldsymbol{\Sigma}_m^{-1}(\tilde{\mathbf{h}}_{m}-\boldsymbol{\mu}_m) d \tilde{\mathbf{h}}_{m} \\ \nonumber
&=  \ln (\pi^n \det(\boldsymbol{\Sigma}_m)) \langle (\tilde{\mathbf{h}}_{m}-\boldsymbol{\mu}_m)^H\boldsymbol{\Sigma}_m^{-1}(\tilde{\mathbf{h}}_{m}-\boldsymbol{\mu}_m)  \rangle  
\stackrel{(b)}{=} \ln (\pi^n \det(\boldsymbol{\Sigma}_m)) \text{trace}(\boldsymbol{\Sigma}_m^{-1}\boldsymbol{\Sigma}_m) \\ \nonumber
&\stackrel{(c)}{\propto} \ln (\det(\boldsymbol{\Sigma}_m))
\end{align}
Equality in (a) is obtained by the definition of expectations. Equaltiy in (b) is obtained by introducing the trace as done in the previous derivation. Proportionality in (c) is obtained by just looking at the terms which are dependent on the variaitonal parameters.
The term $\beta_n$ is trivial to obtain, hence it is omitted.

\textbf{Derivation of Terms of Factorized ELBO:} Here we derive the expression for $\lambda(\boldsymbol{\mu}_m,\boldsymbol{\tau}_m)$ given in equation \eqref{eq:ELBO_factorized}. We have
\begin{align}
\lambda(\boldsymbol{\theta}) = \langle||\tilde{\mathbf{y}}_m-\tilde{\boldsymbol{\Theta}}\tilde{\mathbf{h}}_m||^{2}_2\rangle
\end{align}
The expectation above is w.r.t. the approximating posterior distribution. Since we assume a fully factorized version of the approximating posterior given by \eqref{eq:factorized_posterior}, the variational parameters are given as follows $\boldsymbol{\theta} = \{\boldsymbol{\mu}_m,\boldsymbol{\tau}_m\}$. So
\begin{align}
\lambda(\boldsymbol{\mu}_m,\boldsymbol{\tau}_m)&=\mathbf{\tilde{y}}_m^H\mathbf{\tilde{y}}_m-2\mathbf{\tilde{y}}_m^H\boldsymbol{\tilde{\Theta}}\langle\mathbf{\tilde{h}}_m\rangle + \langle\mathbf{\tilde{h}}_m^H\boldsymbol{\tilde{\Theta}}^H\boldsymbol{\tilde{\Theta}}\mathbf{\tilde{h}}_m\rangle \\ \nonumber
&\stackrel{(a)}{=} \mathbf{\tilde{y}}_m^H\mathbf{\tilde{y}}_m-2\mathbf{\tilde{y}}_m^H\boldsymbol{\tilde{\Theta}}\boldsymbol{\mu}_m + \text{trace}([\boldsymbol{\mu}_m\boldsymbol{\mu}_m^H+\text{diag}(\boldsymbol{\tau}_m)]\boldsymbol{\tilde{\Theta}}^H\boldsymbol{\tilde{\Theta}}) \\ \nonumber
&\stackrel{(b)}{=} ||\tilde{\mathbf{y}}_m-\tilde{\boldsymbol{\Theta}}\boldsymbol{\mu}_m||^{2}_2+\text{trace}(\text{diag}(\boldsymbol{\tau}_m)\boldsymbol{\tilde{\Theta}}^H\boldsymbol{\tilde{\Theta}}) 
= ||\tilde{\mathbf{y}}_m-\tilde{\boldsymbol{\Theta}}\boldsymbol{\mu}_m||^{2}_2+\mathbf{a}^{T}\boldsymbol{\tau}_m.
\end{align}
Equality in $(a)$ is because  $\mathbf{\tilde{h}}_m^H\boldsymbol{\tilde{\Theta}}^H\boldsymbol{\tilde{\Theta}}\mathbf{\tilde{h}}_m$ is a scalar. Equality in $(b)$ is obtained by exchanging expectation and trace and using the cyclic property of trace as stated above. Here $\mathbf{a} = \text{diag}(\boldsymbol{\tilde{\Theta}}^H\boldsymbol{\tilde{\Theta}})$. Rest of the terms for the factorized version are easily derived from the structured version.
\section{}\label{app:A} 
\textbf{Derivation of parameter matrix $\mathbf{A}$:} We maximize the expected value of the complete data log likelihood, by differentiating \eqref{eq:max_surrogate}  with respect to $\alpha_{m}^{NZ}$ and applying first order optimality conditions. The derivative of \eqref{eq:max_surrogate} w.r.t. $\alpha_{m}^{NZ}$ can be written as follows
\begin{align}\label{eq:EM_grad}
\frac{c}{\alpha_{m}^{NZ}}-d+\sum_{n=1}^{NK}\frac{u_{nm}}{{u_{nm}}\alpha_{m}^{\text{NZ}}+{(1-u_{nm})} \alpha_{m}^{\text{Z}}} - \sum_{n=1}^{NK}u_{nm}\mathbb{E}[\tilde{h}_{nm}].
\end{align}
We have to equate this gradient to $0$. We observe that a closed form solution for $\alpha_{m}^{NZ}$ cannot be obtained due to the coupling of the terms. Hence we try to obtain a sub-optimal range of solutions. Since, $\alpha_{m}^{NZ}>0,$ and $\sum_{b=1}^{NK} \beta_{b,n,m} = 1$, we have,
\begin{align}
\label{eq:der3}
\frac{\sum_{n=1}^{NK}u_{nm}}{ (\alpha_{m}^{NZ})^*}>\sum_{n=1}^{NK}\frac{u_{nm}}{{u_{nm}}\alpha_{m}^{\text{NZ}}+{(1-u_{nm})} \alpha_{m}^{\text{Z}}} \geq 0.
\end{align}
After substituting \eqref{eq:der3} in \eqref{eq:EM_grad}, we have
\begin{align}
\label{eq:der4}
\frac{\sum_{n=1}^{NK}u_{nm}}{ (\alpha_{m}^{NZ})^*}>-\frac{c}{(\alpha_{m}^{NZ})^*}+d + \sum_{n=1}^{NK}u_{nm}\mathbb{E}[\tilde{h}_{nm}] \geq 0,
\end{align}
which gives us a range of stationary points given in \eqref{eq:A_range}. Here, we have used the notation $\omega_{nm}=\sum_{b=1}^{NK}\beta_{b,n,m}\mathbb{E}[\tilde{h}_{b,m}]$. Similarly we can obtain the range for $(\alpha_{m}^{Z})^*$.

\vspace{-0.5cm}
\section{}\label{app:convergence}
\textbf{Convergence Proof of FM-SBL:} Here we prove that the updates for the FM-SBL of equations \eqref{eq:varianceFMF} and \eqref{eq:varianceFMF} ascend on the factorized ELBO of \eqref{eq:ELBO_factorized} $\mathcal{L}_{F}$.
\begin{align}
\mathcal{L}_F[\boldsymbol{\mu}_m^{(t+1)},\boldsymbol{\tau}_m^{(t+1)}] &\stackrel{(a)}{=} \mathcal{L}_{FMF}[\boldsymbol{\mu}_m^{(t+1)},\boldsymbol{\tau}_m^{(t+1)};\boldsymbol{\delta}^{(t+1)}] \\ \nonumber
&\stackrel{(b)}{\geq} \mathcal{L}_{FMF}[\boldsymbol{\mu}_m^{(t+1)},\boldsymbol{\tau}_m^{(t+1)};\boldsymbol{\delta}^{(t)}] 
\stackrel{(c)}{\geq} \mathcal{L}_{FMF}[\boldsymbol{\mu}_m^{(t)},\boldsymbol{\tau}_m^{(t)};\boldsymbol{\delta}^{(t)}] 
\stackrel{(d)}{=} \mathcal{L}_F[\boldsymbol{\mu}_m^{(t)},\boldsymbol{\tau}_m^{(t)}]
\end{align}
Equality in $(a)$ is because $\boldsymbol{\delta}^{(t+1)}=\boldsymbol{\mu}^{(t+1)}$. Inequality in $(b)$ is due tothe Lipschitz inequality in  \eqref{eq:lipschitz_inequality}. Inequality in (c) holds due to the gradient ascent on $\mathcal{L}_{FMF}$. Equality in $(d)$ is because $\boldsymbol{\delta}^{(t)}=\boldsymbol{\mu}^{(t)}$. This proves that update equations obtained from optimizing the $\mathcal{L}_{FMF}$, ascend on the factorized objective $\mathcal{L}_F$.

\vspace{-0.1cm}
\bibliographystyle{IEEEtran}
\bibliography{IEEEabrv,references}
\end{document}